%% file: main.tex
\documentclass{aa}
\usepackage[utf8]{inputenc}
\usepackage[T1]{fontenc}
\usepackage{graphicx}
\usepackage{pdflscape}
\usepackage{txfonts}
\usepackage{adjustbox}
\usepackage{geometry}
\usepackage{tabularx}
\usepackage{rotating}
\usepackage{pgffor}
\usepackage[english]{babel}
\usepackage{float} 
\usepackage[normalem]{ulem} 
\usepackage{needspace}
\usepackage{stfloats}
\usepackage{newtxtext,newtxmath}
\usepackage{booktabs}
\usepackage{comment} 
\usepackage[compatibility=false]{caption}
\usepackage{subcaption}
\usepackage{xcolor} 
\usepackage{natbib}
\usepackage{amssymb}
\usepackage[english]{babel}
\usepackage{longtable}
\usepackage{caption}[2020/01/01]
\usepackage[colorlinks=true, linkcolor=blue, urlcolor=blue, filecolor=blue, citecolor=blue,pdfusetitle,pdfdisplaydoctitle,pdfencoding=auto, psdextra]{hyperref}

\definecolor{goldenrod}{rgb}{0.85, 0.65, 0.13}

\definecolor{celine}{rgb}{0.2,0.6,0.8}

\usepackage{hypcap}

\begin{document} 

\title{Gaia FGK Benchmark Stars: Impact of Spectral Resolution on Stellar Abundances}
\author{I. Hern\'andez-Araya \inst{1} \and C. Aguilera-G\'omez\inst{1} \and P. Jofr\'e \inst{2} \and S. Vitali \inst{3} \and L. Casamiquela \inst{4} \and C. Soubiran \inst{5} \and U. Heiter\inst{6} \and S. Blanco-Cuaresma \inst{7, 8} }
    
\institute{Instituto de Astrof\'isica, Pontificia Universidad Cat\'olica de Chile, Av. Vicu\~na Mackenna 4860, 782-0436 Macul, Santiago, Chile \email{ivanna.hernndez@uc.cl} \and Instituto de Estudios Astrof\'isicos, Universidad Diego Portales, Av. Ej\'ercito Libertador 441, Santiago, Chile \and INAF - Osservatorio Astrofisico di Torino, Strada Osservatorio 20, I-10025 Pino Torinese (TO), Italy \and LIRA, Observatoire de Paris, Universit\'e PSL, Sorbonne Universit\'e, Universit\'e Paris Cit\'e, CY Cergy Paris Universit\'e, CNRS, 92190 Meudon, France \and Laboratoire d'Astrophysique de Bordeaux, Univ. Bordeaux, CNRS, B18N, all\'ee Geoffroy Saint-Hilaire, 33615 Pessac, France \and Observational Astrophysics, Department of Physics and Astronomy, Uppsala University, Box 516, SE-751 20 Uppsala, Sweden \and Harvard-Smithsonian Center for Astrophysics, 60 Garden Street, Cambridge, MA 02138, USA \and Faculty of Psychology, UniDistance Suisse, Brig, Switzerland}

\date{Received \today, accepted }
\abstract{In the current era of large Milky Way spectroscopic surveys, calibrating and standardizing stellar parameters and abundance measurements is crucial. The Gaia benchmark stars (GBS) are essential reference points characterized by well-defined parameters derived from fundamental relations independent of spectroscopy.}{We analyze 30 GBS with spectroscopic data at three different resolutions. Our goal is to evaluate the impact of spectral resolution on the measurements of the stellar parameters and chemical abundances. Additionally, we introduce a new set of spectral lines suitable for both metal-poor and metal-rich stars.}{We used  R $\sim$ 190 000 (R190), R $\sim$ 42 000 (R42), R $\sim$ 28 000 (R28), and R190 degraded to R28 (R190-R28) spectral data to measure abundances with synthetic fitting and equivalent widths methods, testing the needed resolution to obtain consistent results.}{Our comparative analysis between R28 and R190-R28 reveals that gaps in wavelength coverage can lead to discrepancies in the derived stellar parameters, particularly in $\log g$. These effects are not primarily driven by spectral resolution, but rather by the limited spectral coverage and line availability. We find overall similar abundance in most cases,  emphasizing the importance of line selection for spectroscopic studies. However, certain elements (e.g, Ti II, Sc II) displayed greater discrepancies, possibly due to the effect of blending that becomes more pronounced as resolution decreases or is HFS-sensitive. Our comparative analysis shows that the abundances for Fe I, Ni I, Ti I, and Si I present less scatter across all resolutions, including R28. }{Our findings indicate that for some elements, the difference between synthetic fitting and the equivalent width method is minimal, especially at the highest resolution. However, we also find that the highest resolution is not always essential for chemical abundance measurements. Our results provide practical guidelines for upcoming large surveys to reconstruct the chemical history of the Milky Way.}


\maketitle

\section{Introduction}

Our understanding of the Milky Way's structure and evolution requires in-depth study and analysis of the stars populating it. In the era of large spectroscopy surveys, various projects have focused on collecting stellar spectra, to characterize stars and eventually reconstruct the Galaxy's formation history by studying the pattern of chemical abundances (\citealp{2017A&A...600A..22M}), gaining insights on the formation site of each different element (\citealp{2019ARA&A..57..571J}). In particular, the Gaia-ESO survey (GES;\citealp{2022A&A...666A.120G}, \citealp{2022A&A...666A.121R}), APOGEE \citep[Apache Point Observatory Galactic Evolution Experiment,][]{2008AN....329.1018A, 2017A&A...600A..22M}, GALAH \citep[Galactic Archaeology with HERMES,][]{2015MNRAS.449.2604D,2025PASA...42...51B}, RAVE \citep[Radial Velocity Experiment,][]{2006AJ....132.1645S}, DESI \citep[Dark Energy Spectroscopic Instrument,][]{2016arXiv161100036D, 2026AJ....171..285D}, SDSS-V \citep[Sloan Digital Sky Survey V, ][]{2007ApJS..172..634A, 2017arXiv171103234K}, WEAVE \citep[William Herschel Telescope Enhanced Area Velocity Explorer,][]{2024MNRAS.530.2688J}, and 4MOST \citep[4-metre Multi-Object Spectroscopic Telescope,][]{2011Msngr.145...14D, 2019Msngr.175....3D} have provided, and will keep providing, a wealth of spectroscopic data including atmospheric parameters (APs) and chemical abundances.

The reliability of the information contained in stellar spectra depends on a variety of aspects (\citealp{2019ARA&A..57..571J}). Different methods are used to extract information from the data. Analyzing how these methods can affect the APs and abundances derived for stars is fundamental to then compare the information with models of galaxy evolution (\citealp{2020A&A...642A.182A}). Comparative studies such as \cite{nandakumar2022}, \cite{2023A&A...670A.107H} and \cite{2022A&A...663A...4S}, which examined data from the GALAH and APOGEE surveys, present clear evidence that differences on common stars are larger than the reported errors.

To improve these differences we need to have common well-studied stars. For this purpose, Gaia FGK benchmark stars (GBS) \citep[][and references therein]{soubiran+2024} play a critical role, offering consistent reference points to align measurements across different spectroscopic studies (\citealp{2025ApJS..281...27P}; \citealp{2019ARA&A..57..571J}), validate pipelines and improve stellar models (e.g., \citealp{2022A&A...666A.120G}; \citealp{2021MNRAS.506..150B}).

The GBS' APs are determined starting from interferometric measurements of angular diameter ($\theta_{LD}$) and bolometric fluxes ($F_{bol}$), used then in fundamental relations, the Stefan-Boltzmann and Newton's law of gravitation, to provide parameters independent of spectroscopy \citep{2015A&A...582A..49H}.
The GBS sample has been assembled to span a wide range of effective temperature ($\mathrm{T_{eff}}$), surface gravity ($\log g$), and metallicities ($\text{[Fe/H]}$), specifically FGK-type stars (including the Sun) and some M-type giant stars \citep{soubiran+2024}.

These stars and their reference properties have been extensively studied, with a growing sample of stars analyzed and published over time. GBS version 1 (GBSv1)  was published in \citet[][Paper I]{2015A&A...582A..49H}, which presents 30 FGK stars and the determination of their $\mathrm{T_{eff}}$ and $\log g$. Subsequently, \citet[][Paper II]{2014A&A...566A..98B} compiled a library of high-resolution spectra, used to determine metallicities \citep[Paper III]{2014A&A...564A.133J} and elemental abundances of $\alpha$-capture and iron-peak elements 
\citep[][Paper IV]{2015A&A...582A..81J}. Later, \citet[][Paper V]{2016A&A...592A..70H} included new candidates with low metallicity and their APs, increasing the sample to 35 stars in a second version called GBSv2 \citep{2018RNAAS...2..152J}.
\citet[Paper VI]{2017A&A...601A..38J} consisted on a careful analysis of systematic uncertainties in abundance determination using different methods, and how the implementation of e.g. normalization in these different methods impacted the resulting abundances. Recently, \citet[][Paper VII]{soubiran+2024} 
released the third version, 
expanding the number of stars to 192 GBS with fundamental $\mathrm{T_{eff}}$ and $\log g$. In addition to the notable increase in the number of stars,  the precision and homogeneity of the new available data $\theta_{LD}$ and $F_{bol}$ determinations improved the parameters, notably for some uncertain ones in GBSv2. Complementary to this, \citet[][Paper VIII]{Casamiquela+2026}, published a new spectral library, determining  chemical abundances for iron-peak and $\alpha-$elements for the GBSv3.

The GBS have become a cornerstone for validating spectroscopic analysis pipelines. Their role is evident in the data products of the Gaia ESO survey \citep{2023A&A...676A.129H}, the GALAH survey, as well as the RAVE survey. Projects such as OCCASO (\citealp{2019MNRAS.490.1821C}) have systematically observed selected  GBS to assess cluster abundances, while upcoming initiatives such as WEAVE and 4MOST are leveraging the GBS for calibration and validation purposes \citep{2026arXiv260218340K}. Also SDSS-V is observing them and using them for validation \citep{2025AJ....170...96M}, and space missions such as Gaia and PLATO (\citealp{2025ExA....59...26R}) include them as validation pillars in their parametrization pipelines. 

This widespread use is possible because the GBS provide a catalogue of accurate astrophysical parameters and chemical abundances, together with a library of high-resolution, high S/N optical spectra. These resources have been widely used to test and advance stellar spectroscopic techniques. For example, \cite{2020A&A...642A.182A} used GBS to compare the ESPRESSO (Echelle SPectrograph for Rocky Exoplanets and Stable Spectroscopic Observations, \citealp{2010SPIE.7735E..0FP}; \citealp{2018haex.bookE.157G}), PEPSI (Potsdam Echelle Polarimetric and Spectroscopic Instrument, \citealp{2015AN....336..324S}), and HARPS (High Accuracy Radial velocity Planet Searcher, \citealp{2015A&A...579A..52N}) spectrographs, ensuring the homogeneity of spectroscopic analyses for cases requiring ultra-high resolution. Similarly, \cite{2022A&A...668A..68A} used GBS to validate 3D non-local thermodynamic equilibrium (NLTE) models, improving the determination of iron abundances and APs by addressing ionization and excitation imbalances, particularly in metal-poor stars. 

In view of existing and upcoming spectroscopic surveys employing instruments with different spectral resolution, we aim to use the GBS to assess how spectral resolution affects derived APs and abundances, and to assess the consistency and precision of these measurements. 
Our analysis thus aims to derive APs and individual abundances of the same lines but different resolutions, focusing on $\alpha$-elements such as Ti, Si, Mg, and Ca; and iron peak elements, including V, Cr, Mn, Co, Ni, and Sc. The $\alpha$ and iron peak elements are crucial for Galactic chemo-dynamical and chemo-kinematical studies (\citealp{2021ApJ...920...23J};  \citealp{2021MNRAS.503.3216P}; \citealp{2021ApJ...913...62F}; \citealp{2020ApJ...901...27S}; \citealp{2020A&A...639A.140S}; \citealp{2019A&A...625A.141L}). %
This paper is organized as follows: Section \ref{Sec:DataMethods} describes the sample data and methodology. Section \ref{Sec:Results} discusses the spectroscopic analysis and presents our main results. In Section \ref{Sec:Discussion}, we see implications for surveys and compare with other line selections, to conclude and summarize in Section \ref{Sec:Summary}.

\section{Data and Methods} \label{Sec:DataMethods}

The stars considered for this study are listed in Table~\ref{Tabla_GBS}. These were taken from the GBSv1 (Paper I), 
and correspond to the 30 southern stars that have been observed by ESPRESSO and analyzed by \cite{2020A&A...642A.182A}.  The APs of these stars used throughout
this work correspond to the values adopted in the GBSv3 Paper VII and VIII. In this section, we describe the data and the methods used to analyze these stars.

\begin{table*}[t]
    \centering
    \setlength{\tabcolsep}{8pt} 
    \renewcommand{\arraystretch}{1.} 
    \footnotesize 
    \begin{tabular}{llllrrrrrrr}
        \toprule
        HIP  &  HD   &  Star   &  Spectral Type  &  $\mathrm{T_{eff}}$   &  $u\_{T_{\mathrm{eff}}}$  &  $\log g$   &  $u\_\log g$   & $\mathrm{[Fe/H]}$ & $u\_\mathrm{[Fe/H]}$ & $v\sin i$
        \\ & &   &  &  (K) & (K)   & (dex) & (dex) & (dex) & (dex) & (km $\mathrm{s^{-1}}$) \\
        \midrule
        HIP79672  &  HD146233   &  18 Sco  &  G2Va  & 5824 & 30 & 4.42 & 0.01 & 0.06  & 0.01& 2.2 \\
        HIP69673    &  HD124897   &    $\alpha$ Boo  &  K1.5III  & 4277 & 23 & 1.58 & 0.07 & -0.55 & 0.04& 3.8 \\
        HIP71681   &  HD128621   &  $\alpha$ Cen B &  K1V  & 5207 & 12 & 4.53 & 0.01 & 0.21 & 0.05 & 1.0 \\
        HIP71683 &  HD128620   &  $\alpha$ Cen A  &  G2V & 5804 & 13 & 4.29 & 0.01 & 0.22 & 0.04  & 1.9 \\
        HIP14135 & HD18884   &  $\alpha$ Cet  &  M1.5IIIa  & 3738 & 170 & 0.66 & 0.07 & -0.56 & 0.17 & 3.0 \\
        HIP37279  &  HD61421   &  $\alpha$ CMi  &  F51V-V & 6582 & 5 & 3.98 & 0.02 & -0.04 & 0.02& 2.8 \\
        HIP21421  &  HD29139   &  $\alpha$ Tau  &  K5III & 3921 & 80 & 1.17 & 0.05 & -0.22 & 0.10 & 5.0 \\
        HIP85258  &  HD157244   &  $\beta$ Ara  & K3Ib-II & 4232 & 17 & 0.97 & 0.06 & -0.04 & 0.23 & 5.4 \\
        HIP37826  &  HD62509   &  $\beta$ Gem  &  K0IIIb  & 4810 & 14 & 2.55 & 0.03 & -0.07 & 0.03  & 2.0 \\
        HIP2021  &  HD2151   &  $\beta$ Hyi  &  G0V & 5917 & 25 & 3.97 & 0.04 & -0.05  & 0.03& 3.3 \\
        HIP57757  &  HD102870  &  $\beta$ Vir &  F9V  & 6093 & 13 & 4.08 & 0.02 & 0.12& 0.02  &  2.0\\
        HIP17378  &  HD23249   &  $\delta$ Eri  &  K1III-IV & 5026 & 38 & 3.83 & 0.10 & 0.03 & 0.04 & 0.7 \\
        HIP16537  &  HD22049   &   $\epsilon$ Eri  &  K2VK: & 5130 & 30 & 4.63 & 0.01 & -0.10 &0.03  &2.4  \\
        HIP63608  &  HD113226   &  $\epsilon$ Vir  &  G8III  & 4950 & 10 & 2.72 & 0.02 & -0.04 &0.03  & 2.0 \\
        HIP67927  &  HD121370   &   $\eta$ Boo  &  G0IV  & 6161 & 18 & 3.82 & 0.05 & 0.32 & 0.04  & 12.7 \\
        HIP98337  &  HD189319   &   $\gamma$ Sge  &  M0III & 3904 & 30 & 1.06 & 0.04 & -0.16 & 0.10& 6  \\
        HIP68594  &  HD122563   &  HD122563  &  F8IV & 4642 & 35 & 1.32 & 0.03 & -2.68 &0.05& 5 \\
        HIP76976  &  HD140283   &  HD140283  &  sdF3  & 5788 & 45 & 3.75 & 0.09 & -2.42 &0.04  & 5 \\
        HIP115227  &  HD220009   &  7 Psc  &  K2III  & 4227 & 77 & 1.66 & 0.04 & -0.73& 0.04& 1 \\
        HIP32851  &  HD49933A   &  HD49933  &  F2V  & 6628 & 89 & 4.19 & 0.04 & -0.44 &0.03  &10  \\
        HIP56343  &  HD100407   &  $\xi$ Hya  &  G7III & 5034 & 34 & 2.78 & 0.07 & -0.009 &0.04 & 2.4 \\
        HIP48455  &  HD85503   &  $\mu$ Leo  & K2III  & 4519 & 23 & 2.43 & 0.06 & 0.20 & 0.06  & 5.1 \\
        HIP8837  &  HD11695   &   $\psi$ Phe  &  M4III  & 3369 & 108 & 0.45 & 0.06 & -1.29 & 0.51 & 3 \\
        HIP8102  &  HD10700   &   $\tau$ Cet  &  G8.5V & 5463 & 16 & 4.52 & 0.01 & -0.46 & 0.02 & 0.4 \\
        HIP14086  &  HD18907   &  $\epsilon$ For  &  K2VFe-1.3CH-0.8 & 5143 & 56 & 3.53 & 0.03 & -0.56 & 0.02  & 4.2 \\
        HIP60172  &  HD107328   &  c Vir  &  K0IIIb  & 4576 & 87 & 1.77 & 0.22 & -0.34 & 0.05& 1.9 \\
        HIP17147  &  HD22879   &  HD22879  &  F9V  & 5962 & 86 & 4.28 & 0.04 & -0.77 & 0.03 &  4.4\\
        HIP48152  &  HD84937   &  HD84937  &  sdF5  & 6484 & 106 & 4.16 & 0.05 & -1.98& 0.07& 5.2 \\
        HIP86796  &  HD160691   &  $\mu$ Ara  &  G3IV-V  & 5974 & 60 & 4.30 & 0.03 & 0.42 &0.04 & 2.2  \\ 
        Sun  &  Sun(Vesta)  & Sun(Vesta)  &  G2V  &  5777  & 1 & 4.44 & 0 & 0.02 &  0& 1.6 \\
        \bottomrule
    \end{tabular}   
        \caption{Gaia benchmark stars with their APs ($\mathrm{T_{eff}}$, $\log g$) determined using the interferometric method (Paper VII), and $\mathrm{[Fe/H]}$ values from Paper VIII, along with their respective uncertainties ($u\_{T_{\mathrm{eff}}}$, $u\_\log g$, $u\_\mathrm{[Fe/H]}$). The $v\sin i$ values were obtained from Paper II. The given uncertainties for the Sun (Vesta) were obtained from Paper I.}
    \label{Tabla_GBS}

\end{table*}

\subsection{Spectroscopic data}\label{star_spectral}

The spectra used in this study were sourced from three distinct databases, each assembled using different instruments whose key difference for this study is their resolution. We call our samples R190, R42 and R28, and we describe them below. {Additionally, we constructed a fourth sample, R190--R28, by degrading the R190 spectra to match the resolution of R28.}

\subsubsection{R190 spectra}

This sample is taken from \cite{2020A&A...642A.182A}, who observed the GBS with the ESPRESSO spectrograph mounted on the Very Large Telescope (VLT; \citealp{2021A&A...645A..96P}; \citealp{2018haex.bookE.157G}) in period P102 and P104 under programs 102.D-0185(A), 0103.D-0118(A), and 0104.D-0362(A).
The data was taken using the ultra-high resolution mode of ESPRESSO, achieving a resolving power of R$\sim$ 190,000 over the wavelength range 380–788 nm. This spectral resolution is approximately four times higher than the nominal resolution of the homogenized spectra presented in Paper II, which were originally resampled to R $\sim$47,000. The ESPRESSO spectra were obtained from the webpage provided by \cite{2020A&A...642A.182A}, where they publish a library of 1D, science-ready spectra with a signal-to-noise ratio (S/N) greater than 100.

\subsubsection{R42 spectra}
The second spectroscopic sample comes from the library of Paper VIII. In short, it includes data from different instruments, namely ESPaDOnS, UVES, HARPS, NARVAL, FIES, ELODIE, and FEROS. The spectra had different resolutions originally, but all have been convolved to a common resolution of $\sim$ 42,000, corresponding to the resolution of the ELODIE spectrograph.
From this library, we selected all available spectra corresponding to the 30 GBS in our sample. In total, we used 113 spectra, reflecting multiple observations of the same star taken with different spectrographs. The spectra were pre-processed by the GBS team as described in Paper VIII, following the methodology of Paper II. The spectra are normalized and corrected for radial velocity and trimmed to the  480 to 680 nm wavelength range.
When multiple observations exist for several stars, the spectra were combined per instrument in \cite{Casamiquela+2026}. These combined spectra, which are publicly available through the CDS and iSpec Website\footnote{\url{https://www.blancocuaresma.com/s/benchmarkstars}}, are the versions adopted in this work. The resulting spectra have S/N values well above 100. In Fig.~\ref{R42-STAR-INSTRUMENT}, we indicate the instrument corresponding to each spectrum used in our analysis.

\subsubsection{R28 spectra}

This sample consists of spectra with a resolution of R $\sim$ 28,000, obtained with the HERMES instrument at the Anglo-Australian Telescope (AAT) in Australia between 2013 November and 2019 February. The spectra were taken as part of the calibration strategy for the GALAH survey \cite{2021MNRAS.506..150B}. Specifically, we used reduced spectra from the GALAH Data Release 3 (DR3) and GALAH DR4, which were downloaded from the publicly available Data Central portal of the Australian Astronomical Observatory (AAO)\footnote{\url{https://datacentral.org.au/services/download/}}.

The R28 spectra consist of one-dimensional, reduced FITS files per star, covering four optical bands, but we use three of them : blue (471–490 nm), green (564–587 nm) and red (647–673 nm). Each FITS file includes several extensions: unnormalised flux (with and without sky subtraction), relative error arrays, and the continuum normalised spectrum. In this case, we used the unnormalised spectrum with sky subtraction to make the homogeneous normalization for the whole sample. The lower SNR for these spectra is 20 in the 6520-6523 Å region; however, the average SNR for the 27 stars is above 100.

This dataset includes all stars of the sample except HD128621 ($\alpha$ Cen B), HD62509 ($\beta$ Gem), and HD102870 ($\beta$ Vir) because they were not observed in GALAH DR3.

\subsubsection{Degraded R190-R28}

In order to disentangle the effects of spectral resolution and wavelength coverage present in the R28 spectra, we constructed an additional dataset by degrading the R190 spectra to a resolving power of R $\sim$ 28,000. The degradation was performed by convolving the original ESPRESSO spectra with a Gaussian kernel corresponding to the target resolution, while preserving their full wavelength range (380–788 nm).
This controlled dataset allows us to isolate the impact of resolution alone, since the observed R28 spectra differ from R190 not only in resolving power but also in spectral coverage. By comparing the degraded R190 spectra with both the original R190 and the observed R28 data, we can separate resolution effects from those introduced by the more limited wavelength coverage in the abundance analysis.

\subsection{Data pre-processing}
For our analysis, we prepared all spectra following a procedure similar to that described in Paper VIII. Since the R42 data were already normalized and corrected for radial velocity (RV), no further processing was applied to them. The same normalization and RV-correction procedure was then applied to the R190 and R28 dataset using \texttt{iSpec} (\citealp{2014A&A...569A.111B}; \citealp{2019MNRAS.486.2075B}). 
\texttt{iSpec} implements two main approaches for deriving APs and abundances: the equivalent-width (EW) method, based on the curve of growth of weak and strong lines, and the synthetic spectral fitting technique (\citealp{2014A&A...569A.111B}). This allows the user to compare and assess systematic differences introduced by the choice of methodology \citep{2019MNRAS.486.2075B}.  The following steps were applied to each spectrum. 

First,  the continuum was defined iteratively by fitting a second-order spline to regions free of strong absorption features. Typically, 100 spline nodes were distributed across the 480–680 nm range, and outlier points deviating more than 2$\sigma$ below the local fit were excluded in successive iterations. 
    The resulting spline function was then used to normalize the observed flux. 
    The same procedure was applied to both datasets (R190 and R28), with the input parameters adjusted to account for their different spectral resolutions.

Then,  the RV shift was determined by applying the cross-correlation algorithm between each observed spectrum and the solar spectrum obtained with the NARVAL spectrograph, which comes with the \texttt{iSpec} package. We follow the same method described in Paper II to correct the RV.

For the particular case of R28 spectra, the observations were present in three separate segments (blue, green and red). These were merged into a single spectrum, introducing gaps of zero flux between the segments.

\begin{figure}[t]
\centering
\includegraphics[width=1.0\columnwidth]{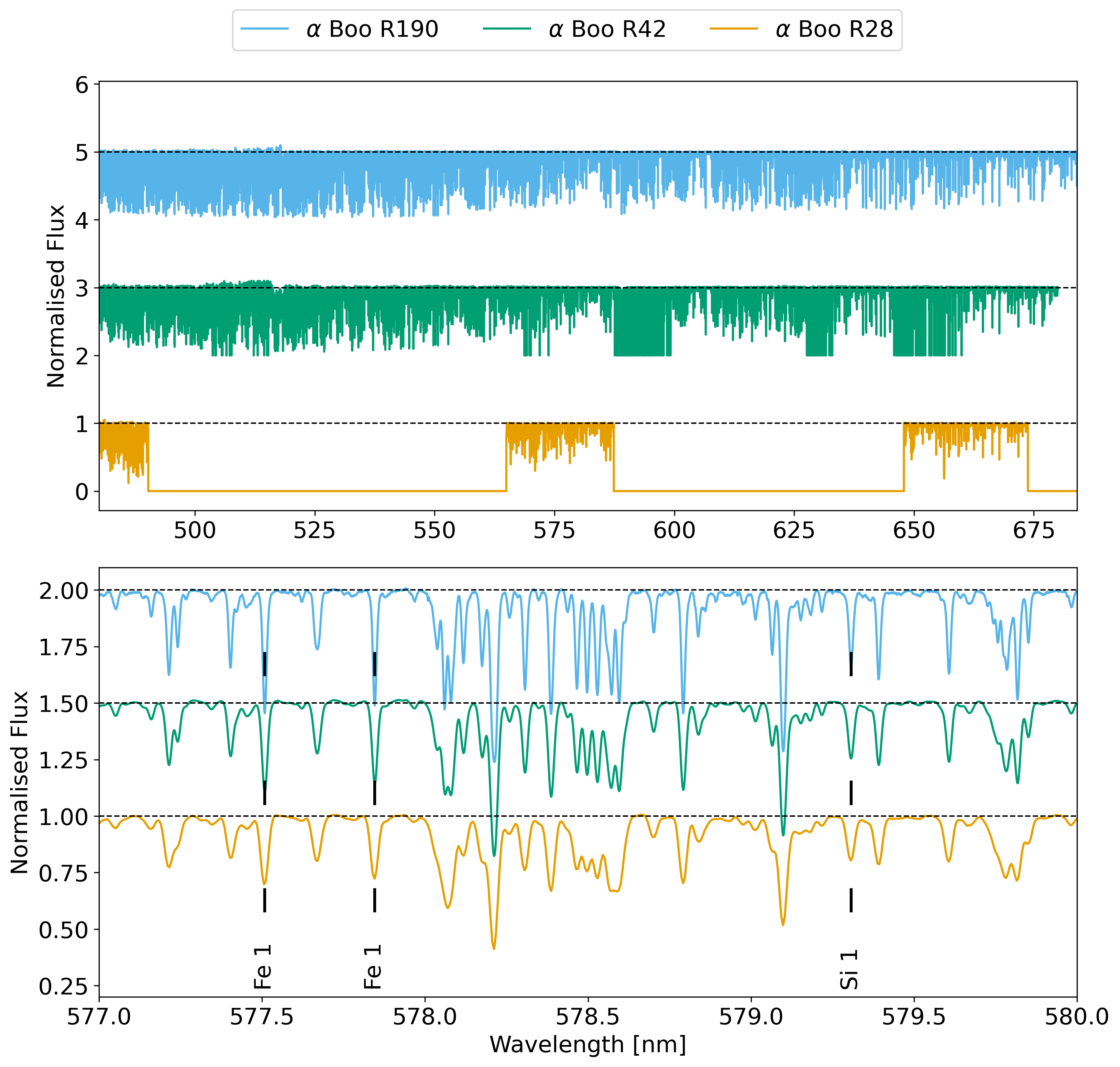}
\caption{Example spectra of $\alpha$ Boo acquired from different instruments with varying spectral resolutions. The top panel shows the spectra on the entire wavelength range, while the bottom panel zooms in on the region around 577-580 nm, highlighting the difference in line shape due to different resolutions.}
\label{alfboo_dif_resolution}
\end{figure}

An example of the processed data is shown in Fig.~\ref{alfboo_dif_resolution}, where we illustrate the spectrum of $\alpha$ Boo. The top panel shows the normalized and RV corrected spectra of the entire wavelength coverage considered in this work for each dataset. The bottom panel shows a zoom-in region of the spectra in a range of 3~nm.

As spectral resolution increases, weak absorption lines become progressively better defined, while strong lines, such as Fe I and Si II, labeled in the figure, are well resolved in all spectra. We can further see that the R28 spectra have a significantly reduced wavelength coverage compared to R42 and R190, with large gaps between spectral segments. These discontinuities result in the loss of important absorption features, limiting the number of lines that can be measured and compared.

\subsection{Line Mask}\label{spectral_line}

As the basis for our line selection we use the line list of \cite{2021A&A...645A.106H}, which was developed for the Gaia-ESO survey, hereafter referred to as the GES linelist. It originates from VALD (\citealp{1995A&AS..112..525P}; \citealp{2015PhyS...90e4005R}), but incorporates updated atomic data for several transitions \citep{2021A&A...645A.106H}. This compilation aimed to identify the most reliable lines for each chemical element by assessing atomic data quality and blending characteristics. To do so, \cite{2021A&A...645A.106H} synthesized a preselected set of lines for the Sun and Arcturus at R = 47,000 and inspected a list of observed GBS spectra; these were used to guide both the line selection and flagging system \citep[see Appendix B of][]{2021A&A...645A.106H}.

This linelist, which includes hyper-fine structure (HFS) splitting where appropriate, comprises 2631 lines between 475 and 686 nm. Two diagnostic flags describe line quality: {\tt syn\_flag} for blending behavior and {\tt gf\_flag} for atomic data reliability.
{\tt syn\_flag} assesses blending by comparing synthetic and observed spectra. It takes values of \textbf{``Y''} (unblended or blended only with the same species in both Arcturus and the Sun), \textbf{``U''} (uncertain blend properties); or \textbf{``N''} (too blended to be useful). On the other hand, {\tt gf\_flag} reflects the quality of the transition probabilities, that is, the $gf$-values. It takes values of \textbf{``Y''} (high accuracy), \textbf{``U''} (inconclusive), or \textbf{``N''} (low accuracy).

Because the original flagging system was developed for spectra at R = 47,000, a key objective of our work was to test whether lines flagged as {\tt syn\_flag}=U in the GES linelist could be reclassified as Y at the higher resolution of our R190 spectra. We therefore selected the 1216 flagged lines with {\tt syn\_flag} = Y and U, and {\tt gf\_flag} = Y from the GES linelist and evaluated them in 13 R190 spectra. These 13 stars were chosen to span the full range of APs covered by the complete sample. As a result, they probe the diversity of blending conditions and line strengths expected across the dataset. While the same inspection could in principle be extended to all stars in the sample, the selected subset already samples the full parameter space, and additional stars would not alter the line classification but only repeat similar assessments under similar atmospheric conditions. The stars analyzed were: $\alpha$Boo, $\alpha$Cet, $\alpha$Cmi, $\beta$Ara, $\beta$Hyi, $\delta$Eri, $\epsilon$Eri, $\gamma$Sge, 7 Psc, $\tau$Cet, $\epsilon$For, HD22879, and $\mu$Ara, spanning $T_{\mathrm{eff}} \sim 3700$--$6500\,\mathrm{K}$, $\log g \sim 0.66$--$4.63\,\mathrm{dex}$, and $\mathrm{[Fe/H]} \sim -0.77$ to $0.42$.

 We considered a line suitable for analysis if it satisfied all of the following conditions:
\begin{enumerate}
    \item It is not blended with any other line, including those of the same species. Blending was assessed by comparing the observed spectrum with synthetic spectra computed (i) excluding the target transition and (ii) including the full line list, in order to identify additional contributors in the surrounding wavelength region.
    \item Its normalized flux at line center is not lower than 0.2, avoiding saturation.
    \item It is clearly distinguishable from the continuum noise, with a detection greater than 1.5$\sigma$.
    \item Its profile is consistent with a simple analytical shape (Gaussian, Lorentzian, or Voigt). Lines showing asymmetries, extended wings, or irregularities inconsistent with these profiles were discarded, and their shapes were validated through Gaussian fitting.
\end{enumerate}

\begin{figure}[t]
\centering
\includegraphics[width=\columnwidth]{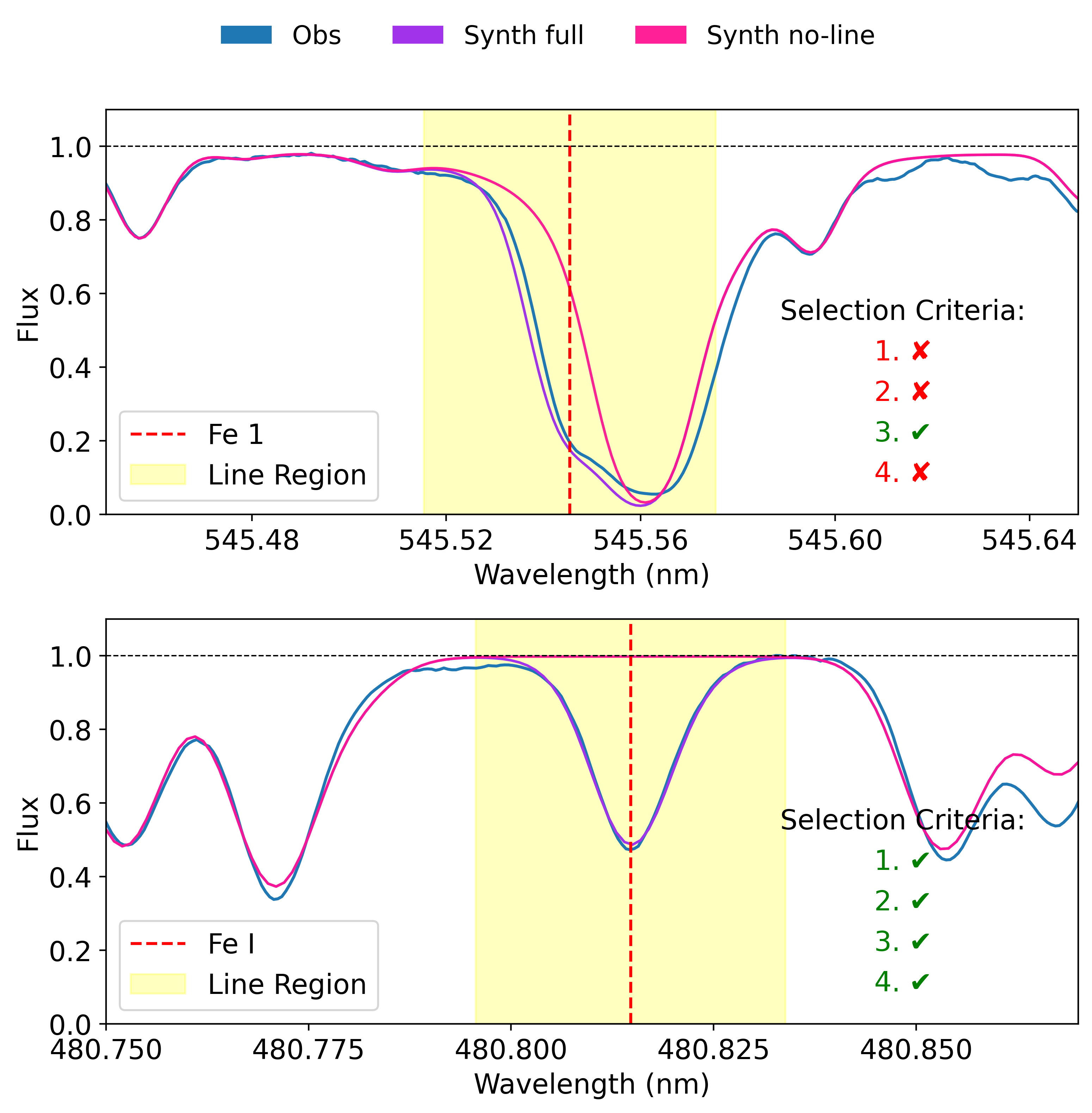}
\caption{Spectrum of $\alpha$ boo (Arcturus) at R190. The upper panel shows a spectral line that does not fulfill the selection criteria, in particular criteria 1, 2, and 4, while the lower panel shows a line classified as suitable. The solid purple curve represents the synthetic spectrum at R190 computed with the full line list, while the dotted pink curve shows the synthetic spectrum at R190 computed after removing the transition under evaluation.}
\label{alfboo_galah_good_bad}
\end{figure}

Figure~\ref{alfboo_galah_good_bad} illustrates two Fe I lines in Arcturus at R190: a rejected line at 545.54 nm (upper panel, with {\tt syn\_flag}=Y) and a selected line at 495.01 nm (lower panel with {\tt syn\_flag}=U). 
The former is blended and saturated, with a non-Gaussian profile, while the latter fulfills all selection criteria. To assess blending (criterion 1), we compared synthetic spectra computed with the full line list and with the target transition removed. If removing the transition causes the absorption feature to vanish and no nearby residual absorption remains, this indicates that the line is not dominated by other transitions and can be safely used. Conversely, if significant absorption persists after removing the transition, the feature is dominated by blends and is therefore rejected.

Once the evaluation was completed, we analyzed the presence of each line across all 13 spectra. A line was included in the master selected line list if at least 80\% of the spectra classified it as a good line. This threshold was adopted because the two cooler stars in the sample were particularly complex and did not meet the four conditions for most lines. 
By setting the threshold at 80\%, we were able to retain lines that performed well in the majority of cases without being overly restrictive. This approach ensured that only lines with consistent, high-quality ratings across most spectra were retained in the final list. Our analysis showed that we need two separate line selections depending on stellar metallicity, separated at $\mathrm{[Fe/H]} = -1.5$~dex. 

To quantify the impact of our re-evaluation, we compared the original GES classifications with our updated flags. For metal-rich stars, 218 lines were reclassified from {\tt syn\_flag}=Y to N, while 46 lines improved from U to Y. For metal-poor stars, 288 lines were downgraded from Y to N, and 148 lines were upgraded from U to Y.

Our master-selected line lists are made available online. Each line in this list is annotated with its corresponding element and peak wavelength. Additionally, for each line, we specify the region of interest, detailing the lower and upper limits in wavelength of the lines.

\subsection{Spectroscopic analysis}

Using \texttt{\texttt{iSpec}} (version 20180608), APs are measured with spectral synthesis and abundances are measured with both synthesis and the EW method. For both analyses, we use the radiative transfer code MOOG (\citealp{1973PhDT.......180S}), which is designed for analyzing lines within a one-dimensional  plane-parallel, and LTE framework and consider 1D LTE MARCS atmospheric models (\citealp{2008A&A...486..951G}). The atomic data were sourced from the GES linelist, and the solar abundances considered are those of \cite{2007SSRv..130..105G}. Although the recommended abundances for the GBSv3 were obtained using TURBOSPECTRUM, we adopt MOOG here for methodological consistency and computational efficiency. In particular, MOOG enables a homogeneous comparison between abundances derived from spectral synthesis and from EW measurements within the same radiative-transfer framework, which is a primary goal of this work. In order to guarantee reliable measurements, only lines with reduced equivalent widths (REW) in the range -6.7 < REW < -4.5 are considered when the EW method is applied, following Paper VIII. The implementation of TURBOSPECTRUM in \texttt{\texttt{iSpec}} is primarily optimized for spectral synthesis analysis, especially when including extensive molecular line lists, whereas our study focuses on differential comparisons between methods and spectral resolutions. We emphasize that this work does not aim to redefine the recommended GBS abundance scale, which remains anchored to the TURBOSPECTRUM results for the reasons extensively explained in Paper VIII.

\begin{figure*}[t]
    \centering \includegraphics[width=0.85\linewidth]{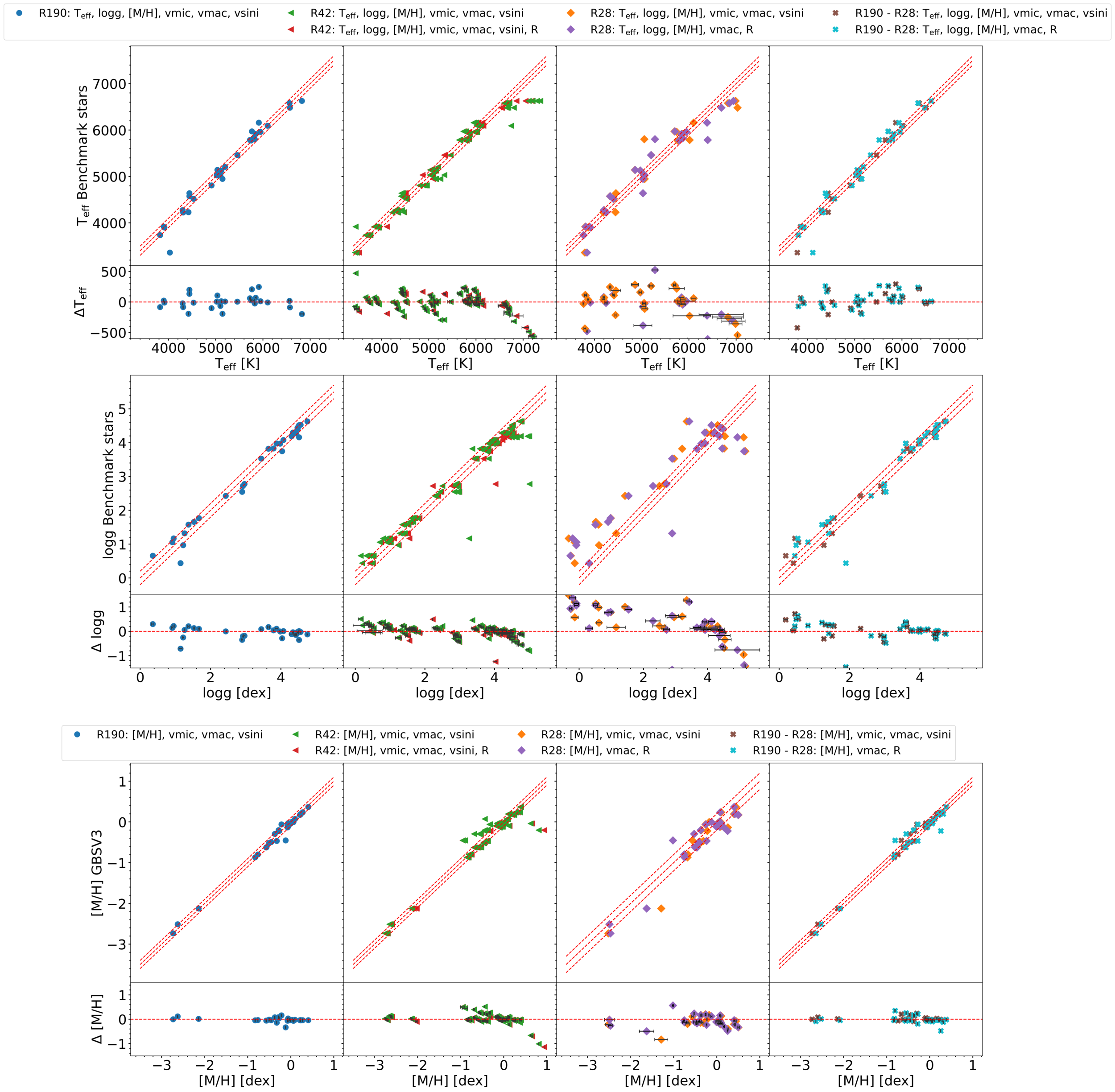}  \caption{Comparison of APs ($\mathrm{T_{eff}}$, $\log g$, $\mathrm{[M/H]}$) with reference values  for the same combination of free parameters used in \texttt{iSpec} at resolution R190 (blue; left columns), R42 (green; second columns),  R28 (orange; third columns) and R190 degraded at R28 (brown; fourth columns). The red triangle, purple diamond, and cyan asterisk symbols in their respective panels represent the best combination of free stellar parameters for R42, R28, and R190-R28. Dashed red lines indicate the 1:1 correlation. For $\mathrm{T_{eff}}$, we plotted $\pm$100 K, while for $\log g$ and $\mathrm{[M/H]}$, we plotted $\pm$0.2 dex. The residuals ($\Delta T_{\mathrm{eff}}$, $\Delta \log g$, $\Delta \mathrm{[M/H]}$) are shown in the bottom row of each subplot, highlighting the discrepancies between the measured temperatures and the benchmark values. Mean offsets and dispersions can be found in Table~\ref{tabla:std_mean}. }
     \label{fig: teff_logg_feh_all}
\end{figure*}

\section{Results} \label{Sec:Results}
\subsection{Stellar Parameters}\label{stellar_parameters_data}

To test and validate our R190-based line selection, we measured APs for the sample stars. With this carefully crafted line selection, we expect to recover the fundamental GBSv3 APs at all resolutions.  

A second goal of this section is to evaluate how resolution affects AP determination purely based on spectroscopy. We derived $\mathrm{T_{eff}}$, $\log g$ and  $\mathrm{[M/H]}$\footnote{$\mathrm{[M/H]}$ represents the global metallicity. Although $\mathrm{[M/H]}$ is correlated with the iron abundance $\mathrm{[Fe/H]}$, the two quantities are not equivalent}. We then compared our result with the GBSv3 reference values listed in Table \ref{Tabla_GBS} to assess their accuracy. We determined APs using spectral synthesis, varying iteratively the parameters to be determined until the best fit to each spectrum was obtained. \texttt{iSpec} can fit various parameters simultaneously. These are $\mathrm{T_{eff}}$, $\log g$, $\mathrm{[M/H]}$, $\alpha$-enhancement, microturbulence velocity $v_\mathrm{mic}$, macroturbulence velocity $v_\mathrm{mac}$, rotational velocity $v\sin i$, and spectral resolution. A subset of these parameters were fitted in our analysis, while others are left fixed, as detailed below.

Limb darkening, i.e., the decrease in specific intensity from the center of the stellar disk toward the limb, was treated using a linear limb-darkening law. In this parametrization, the intensity is described by a single coefficient which quantifies the strength of the center-to-limb variation. Following the implementation in \texttt{iSpec} and adopting the same approach as \cite{2019MNRAS.486.2075B}, we fixed the coefficient to $u = 0.6$, which is representative of FGK-type stars in the optical regime. The limb-darkening coefficient enters the rotational broadening kernel (\citealp{2008oasp.book.....G}) and primarily affects the detailed shape of rotationally broadened profiles. Since our targets are slow rotators, fixing this parameter does not introduce significant systematic effects in the derived APs. RVs were fixed to zero after applying the corresponding corrections during the spectral pre-processing. Initial APs were adopted from GBSv3 and from the literature values compiled in Paper II and used as starting guesses in the fitting procedure. The instrumental resolution was fixed to the nominal value of each dataset when it was not treated as a free parameter as explained below.

To evaluate how APs are recovered at different spectral resolutions, we implemented a two-step fitting strategy.
In the first step, we determined the optimal configuration of free parameters at each resolution. For the R190 spectra, we found that the optimal set of free parameters, providing the best agreement with the GBSv3 reference APs is given by: $\mathrm{T_{eff}}$, $\log g$, $\mathrm{[M/H]}$, and the three broadening parameters $v_\mathrm{mic}$, $v_\mathrm{mac}$, and $v\sin i$.

For the lower-resolution datasets (R42, R28, and the degraded R190–R28 spectra), we additionally allowed the spectral resolution to vary as a free parameter within \texttt{iSpec}. This accounts for possible differences between the nominal instrumental resolution and the effective line broadening in the spectra. For each resolution, we tested to fit $\mathrm{T_{eff}}$, $\log g$, $\mathrm{[M/H]}$, $R$, and all or a subset of the three broadening parameters.
We identified the parameter configuration that minimized the offsets relative to the GBSv3 reference values, defining a preferred set of free parameters for that resolution.
In the case of R42 the preferred set includes all of the three broadening parameters, while for the R28 and R190–R28 datasets only one of the three broadening parameters was included, namely $v_\mathrm{mac}$.

In the second step, after verifying that $\mathrm{T_{eff}}$ and $\log g$ could be consistently recovered across resolutions, we fixed them to the GBSv3 values. We then allowed $\mathrm{[M/H]}$, $v_\mathrm{mic}$, $v_\mathrm{mac}$, and $v\sin i$ to vary, alternatively complemented by $R$ and the same subset of the three broadening parameters as above. This allowed us to refine the metallicity determination under consistent APs.

Figure \ref{fig: teff_logg_feh_all} compares derived $\mathrm{T_{eff}}$, $\log g$, and $\mathrm{[M/H]}$ at  R190, R42, R28, and R190-R28. Mean offsets and dispersions relative to GBSv3 are listed in Appendix~\ref{app: APs}. The two challenging cool giants ($\psi$~Phe and $\beta$~Ara) were excluded from these statistics to maintain homogeneity across resolutions, since they exhibited large discrepancies in the R190 data due to spectral complexity and strong line blending.

We see in Fig.~\ref{fig: teff_logg_feh_all} that GBSv3 APs are recovered for most stars at all resolutions,  with mean offsets increasing as resolution decreases. The values for the offsets can be found in Table~\ref{tabla:std_mean}. Using the R190-optimized free parameter set ($\mathrm{T_{eff}}$, $\log g$, $\mathrm{[M/H]}$, $v_\mathrm{mic}$, $v_\mathrm{mac}$, and $v\sin i$) at lower resolutions yields results comparable to fitting each dataset independently, so we adopted it for all further analyses including the AP comparisons and the abundance determination. Dispersion grows at R28 probably due to reduced wavelength coverage, but APs remain broadly consistent. Regarding R42, the standard deviation can be lower since some stars have multiple observations.

Outliers primarily correspond to $\psi$~Phe, $\beta$~Ara, as well as isolated cases where observational conditions affected individual spectra ($\gamma$~Sge, $\xi$~Hya). These discrepancies reflect the difficulty of modeling molecular features with the current code. This was one of the reasons of why for the spectroscopic analysis of the GBSv3 the recommended results come from TURBOSPECTRUM (see discussions in Paper VIII). In R28, blending between lines could be affecting the determination of APs such as those of $\alpha$~Cen~A: We measure a temperature of $5177 \pm 45 $ K, which is consistent with GALAH's reported temperature of 5271 K (\citealp{2018MNRAS.478.4513B}), but significantly lower than the reference of $5804$ K. At higher temperatures, stars such as $\alpha$ CMi and HD49933 also show important discrepancies with $\Delta T_{\mathrm{eff}}$ of 230 K and 369 K, respectively.

Figure~\ref{fig: teff_logg_feh_all} shows an increased dispersion and offset in $\log g$ at R28. To investigate whether this effect is primarily driven by spectral resolution or by the loss of diagnostic lines, we compare with the degraded the R190-R28, repeating the parameter determination. The resulting dispersion is comparable to that obtained from the native R190 spectra, indicating that the dominant effect of the offset is due to the reduced number of $\log g$ sensitive lines in these spectra. 
At R28, the presence of wavelength gaps removes entire spectral regions containing useful Fe\,I and Fe\,II lines. As a result, fewer diagnostic lines are available to constrain $\log g$, increasing degeneracies between APs and producing the observed offset. We note that the GALAH data releases use more information to derive $\log g$ which come from Gaia parallaxes \citep{2025PASA...42...51B}. 

The third row in Fig.~\ref{fig: teff_logg_feh_all} compares our measured $\mathrm{[M/H]}$ with Paper VIII values. 
At R190, cool giants such as  $\psi$~Phe, show difference in metallicities reflecting the difficulty of isolating unblended lines in these complex spectra (see an example in Appendix \ref{lines}). Other stars that show discrepancies in metallicity at lower resolutions are $\alpha$~Tau, $\xi$~Hya, and HD~84937. Most other stars show good agreement, indicating that our line selection and fitting approach recovers reliable metallicities across resolutions except for the most challenging targets. 

We obtain a small increase in scatter at lower resolution, but the APs are still accurate. The largest discprepancy is seen at R28 but this could be rather a wavelength coverage issue and not a resolution one because for R190-R28 the results are similar to the other ones.  
The good agreement with the GBSv3 reference scale confirms the reliability of our spectroscopic setup and line selection. We therefore adopt the GBSv3 APs (Paper VII) for all subsequent abundance analyses.

\subsection{Elemental abundances}\label{abundances method}

In this section we present our results on abundances. The APs adopted are those indicated in Table~\ref{Tabla_GBS}. Abundances are derived using the {\tt iSpec} routine either by fitting all selected lines simultaneously (which we refer to as global abundances) or by performing a line-by-line abundance determination to identify outliers and assess consistency.

In Table~\ref{tabla:elements} we indicate the number of lines used to measure each elemental abundance, and where hyperfine splitting  (HFS) has to be considered. 
Although Mg lines (the only three lines available, i.e., 516.73, 517.26, and 518.36 nm ) did not meet all quality criteria discussed in Sect.~\ref{spectral_line}, we included this element in our analysis due to its popularity as tracer of $\alpha$ elements. At R28, spectral gaps prevent analysis of certain elements entirely.

\begin{figure*}[t]
\centering
\includegraphics[width=0.85\textwidth]{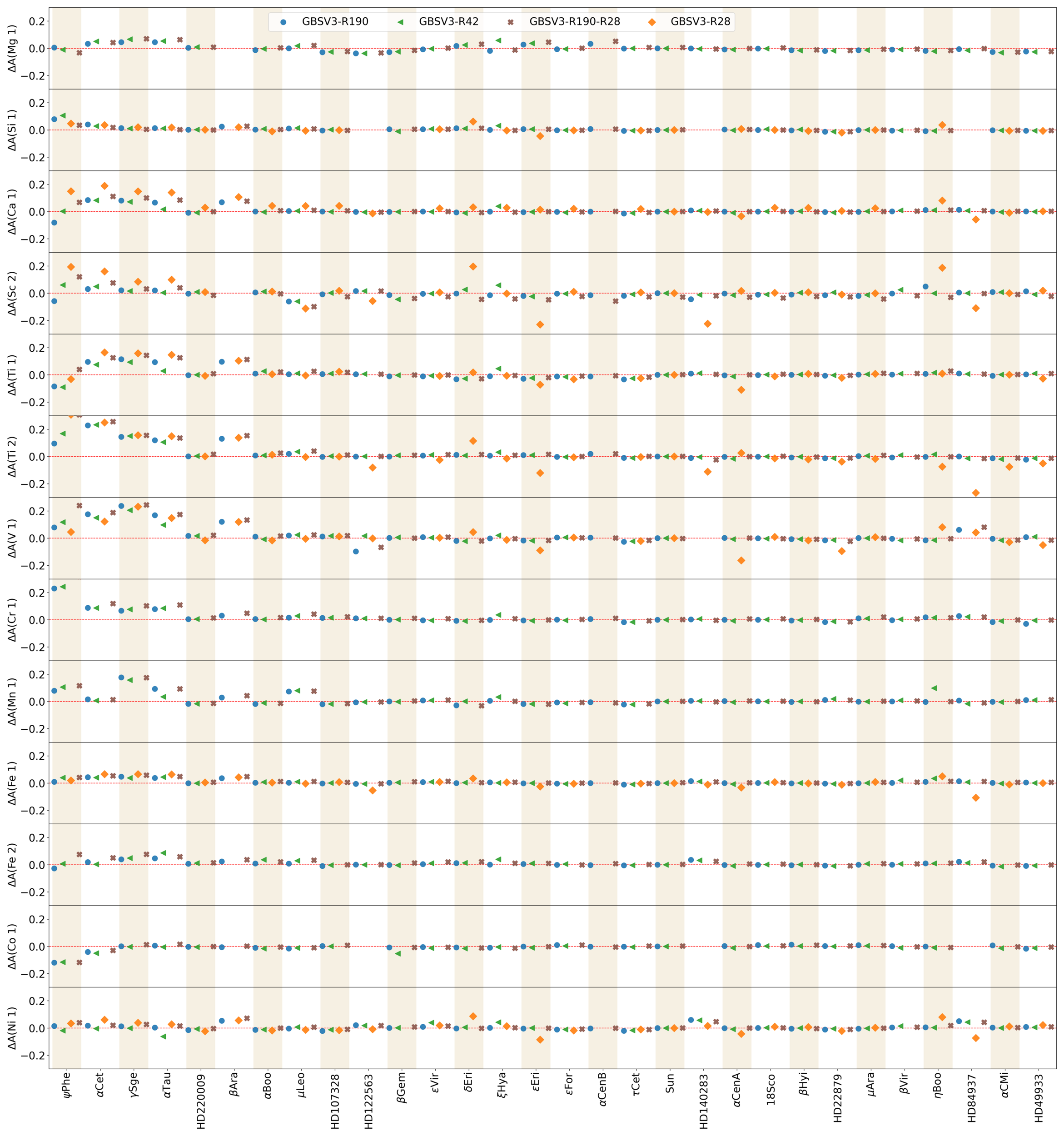}
\caption{Global abundance differences of all elements (top to bottom panels) between the MOOG results of Paper VIII and our results derived for R190, R42, R28, and R190-R28. The GBS sample is ordered by effective temperature, with cooler stars on the left and hotter stars on the right. }
\label{GBSV3-R190}
\end{figure*}

\subsubsection{Comparison with reference abundances}\label{Comparison with reference abundances}

We compare the abundances determined in this work with those from GBSv3 (Paper VIII) obtained using MOOG. The comparison was conducted using the same atmosphere models and solar abundance scale adopted in that paper.  Results for all resolutions are shown in Fig.~\ref{GBSV3-R190} and summarized in Table \ref{tab:mean_abs_delta_gbsv3_fmt}. The stars are sorted by effective temperature, and each resolution is represented with a different color and symbol. Individual stars are indicated by shaded vertical bands in all panels. The abundance differences span the same range in all cases, spanning from -0.25 to 0.25 dex.

Overall, our abundances show excellent agreement with the MOOG results of Paper VIII, particularly for the solar-type stars, supporting our line selection. Towards cooler temperatures, differences increase for all resolutions, in some cases, such as Ti and V, exceeding 0.2 dex, mainly due to stronger spectral blending. Examples of blended features affecting Ca\,I, Ti\,I, and Ti\,II lines are presented in Appendix~\ref{lines}. Metal-poor stars also tend to show larger discrepancies, likely due to the weaker line strength, that become difficult to detect at lower resolution. In our global determination of abundances, we fit the abundance to all lines, regardless of their depth. 

Although no clear dependence on resolution is visible in Fig.~\ref{GBSV3-R190}, the mean differences reported in Table~\ref{tab:mean_abs_delta_gbsv3_fmt} reveal subtle trends. We find excellent agreement for Mg\,I, Si\,I, Fe\,I, Fe\,II, Co\,I, and Ni\,I at all resolutions, with the lowest dispersion at R190 ($\sim$0.01–0.02 dex). 

A larger scatter is observed for Ca\,I, Sc\,II, Ti, V\,I, Cr\,I, and Mn\,I, in cool stars. In most cases, R190 and R42 yield comparable results. For example, for $\gamma$~Sge the mean difference for Sc II is 0.022 at R190 and 0.016 at R42. In contrast, R28 shows increased dispersion, particularly for Ti\,II (0.030 $\pm$ 0.055 dex at R190 versus 0.077 $\pm$ 0.087 dex at R28), reflecting the combined effects of blending and reduced line availability.

The behavior observed in cool stars has also been reported in Paper~VIII, where the largest discrepancies with the literature are also found for Ti, V, and Mn. This is particularly notable in the coolest stars of the sample ($T_{\mathrm{eff}} \leq$ 4000-4500 K), where we observed the same trend in Fig.~\ref{GBSV3-R190}. Magnesium is also known to be a delicate element, as in many cases only a single usable line is available. In addition, Paper~VIII has shown that its abundance can be overestimated depending on the radiative transfer code, with more consistent results obtained using TURBOSPECTRUM. Although we find overall good agreement with the results from Paper VIII, our values may still be slightly overestimated.  We find no significant differences between R190 and R42, but more outliers appear at R28 and not in R190-R28, suggesting that the source of the difference is not the resolution but the number of available lines. The outliers become more prominent for the metal-poor stars.

In summary, spectral resolution has a limited impact on the derived abundances. Most elements show consistent results across R190, R42, and R28, and the increased scatter at R28 is primarily driven by line-selection effects associated with its spectral coverage rather than by resolution itself.

\begin{figure*}[t]
\centering
\includegraphics[width=0.85\textwidth]{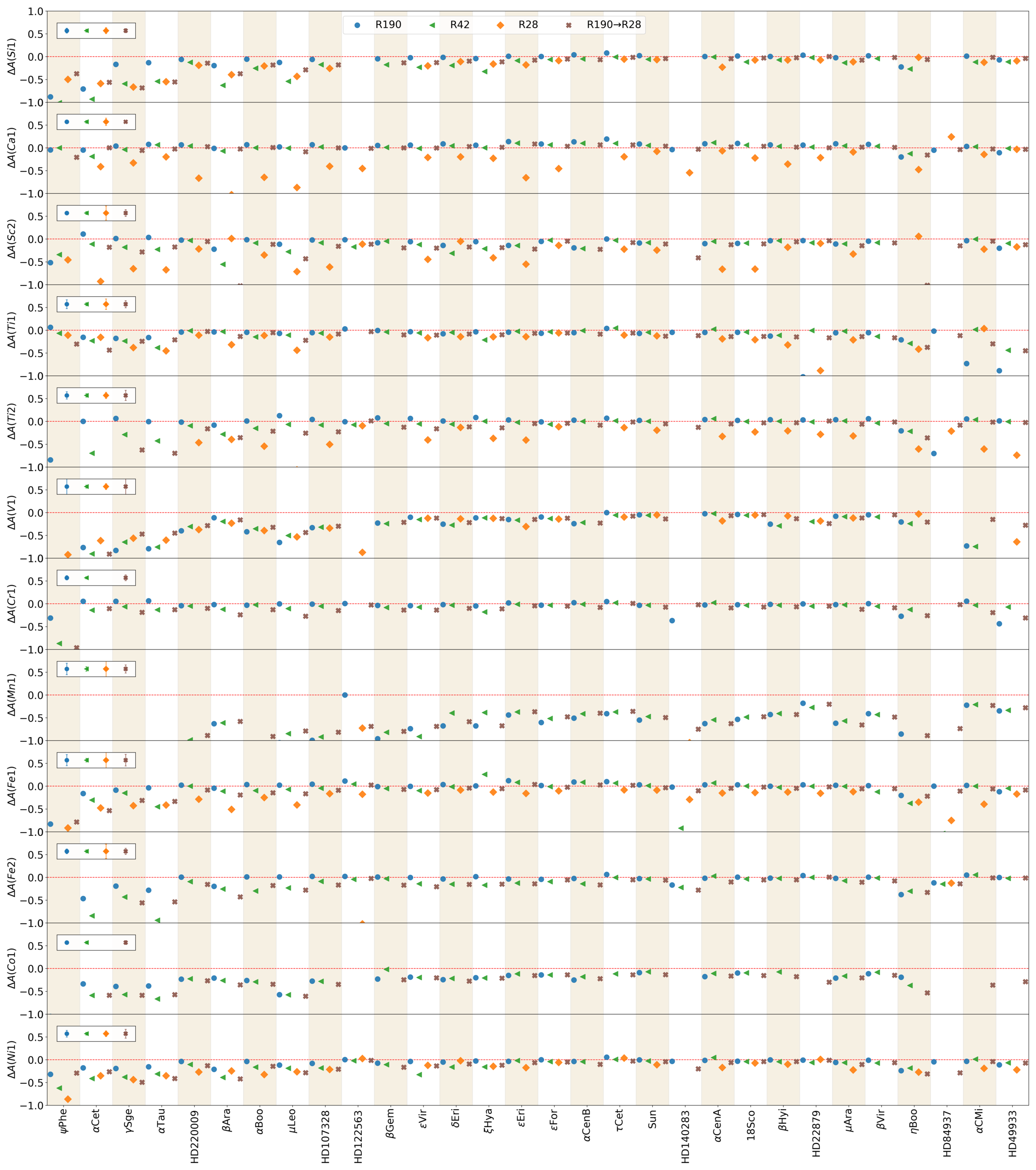}
    \caption{Global abundance differences ($\Delta  A\left( X \right)$) between the synthetic fitting method and EW for all the elements (top to bottom panels) for each star ordered by $\mathrm{T_{eff}}$.
    }
    \label{fig: abun_1}
\end{figure*}

\subsubsection{Comparing abundances derived with Synthesis and EWs} \label{synth_ew_results}

We now compare abundances derived with the synthesis and EW approaches, adopting the same atmospheric models, radiative transfer code, and line selection (Section \ref{spectral_line}). Gaussian profiles were fitted to compute EWs \citep{2014A&A...569A.111B}. If lines are clean and unblended, both methods are expected to return similar abundances, regardless of resolution.

Figure \ref{fig: abun_1} shows abundance differences between synthesis and EWs ($\Delta A(X) = A(X)_\mathrm{synth} - A(X)_\mathrm{EW}$) for the same elements and in the same color scheme as in Fig.~\ref{GBSV3-R190}. Mean differences and dispersion for each element can also be found in Table \ref{tab:mean_abs_synth_minus_ew_fmt}. Errors combine the uncertainties of both methods.

Overall, dispersion increases toward cooler stars, with EW measurements tending to overestimate abundances. Indeed, for $\psi$~Phe differences can reach $\sim$1 dex. This result is likely related to the use of Gaussian profiles used to fit spectral lines, which may not adequately reproduce blended and crowded features that become increasingly important in cool stellar atmospheres. A clear increase in scatter is observed at R28, as shown in Table \ref{tab:mean_abs_synth_minus_ew_fmt}. For example, Ca I shows a mean difference of 0.4~dex at this resolution. In contrast, the degraded R190-R28 spectra present a very small  difference of 0.04~dex, highlighting the importance of both high-quality line selection and spectral coverage over resolution to obtain precise abundance determinations

Several elements show good agreement between the two methods for R190 (e.g., Ti I, Ti II, Cr I, Fe I, Fe II, Ni I), particularly near solar temperature, typically within $\lesssim$ 0.15 dex. However, these differences have a dependence with resolution.  While offsets at R190 are small, they  increase for R190–R28, and R42. For instance, Ni\,I shows mean differences of 0.075 dex at R190, 0.206 dex at R42, and 0.156 dex for R190–R28. 
 
We excluded Mg\, I abundances from this analysis, since the lines used are strong and thus not suitable for the comparison between methods. For Ca\,I, and to some extend Fe II and Ti II,  discrepancies at R28 are primarily driven by the limited number of suitable lines in this spectral range.

Elements affected by HFS, such as V\,I, Mn\,I, and Co\,I, show systematic differences between the two methods. In the case of V\,I, a clear trend with effective temperature is observed, with larger discrepancies toward cooler stars. While synthesis naturally accounts for HFS, the EW method treats each feature as a single line, which can lead to overestimated abundances. This is particularly evident for V I, where we find mean differences of 0.62 dex at R190, 0.38 dex at R42, 0.32 dex at R28, and 0.46 dex for the R190–R28 comparison, clearly indicating systematic overabundances when using the EW method. We note that our cut in reduced EW is applied only for the abundances determined with EWs, therefore this comparison does not consider exactly the same lines. This might trigger further discrepancies in the final result, explaining why the differences here might be larger than those studied in e.g. Paper IV and Paper VI. Further tests on line selection will be found in the next section. 

Although EW measurements themselves do not incorporate HFS, most studies correct for it at the abundance-determination stage. A common approach, adopted recently in \cite{2024A&A...689A.201S}, is to measure EWs with automated tools such as DAOSPEC (\citealp{2008PASP..120.1332S}), and then apply HFS corrections in MOOG using the blends or synth options together with linelists from Linemake (\citealp{2021RNAAS...5...92P}) that provide the individual HFS components. For species without HFS (e.g. Ca, Fe), abundances are simply derived with the abfind option, while for elements affected by HFS, MOOG accounts for the HFS splitting at this stage.

In summary, spectral synthesis and EW methods yield consistent results for well-behaved lines and warm stars, but differences increase toward lower temperatures and lower spectral coverage and for elements affected by HFS. These discrepancies indicate that spectral synthesis is generally more robust, particularly for complex or blended features.

\begin{figure*}[t]
    \centering
    \includegraphics[width=0.85\textwidth]{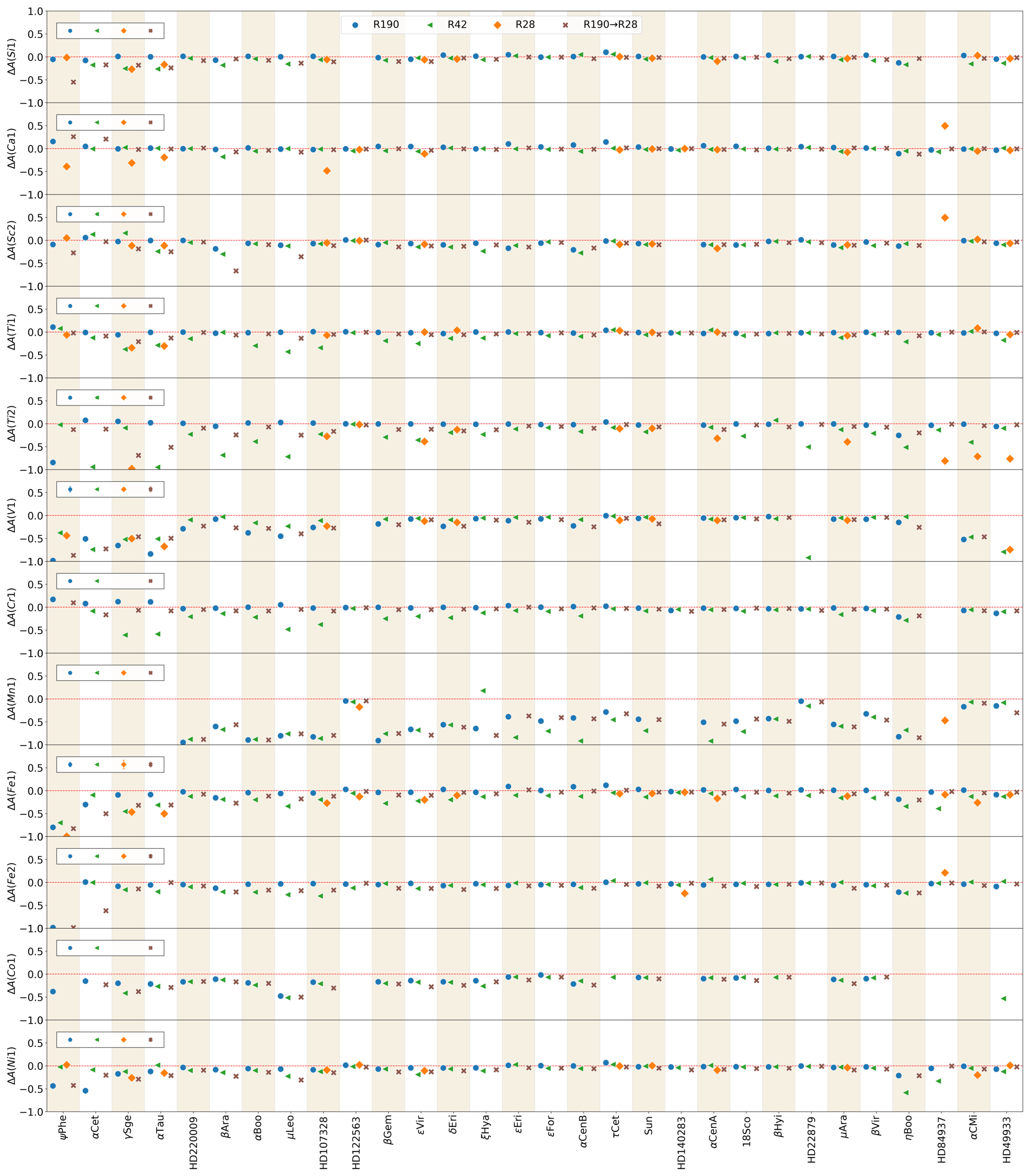}
    \caption{Abundance differences ($\Delta  A\left(Synth-EW \right)$) line-by-line between the synthetic fitting method and EWs across different elements per stars (top to bottom panel). 
    The error bars represent the standard deviation among different lines for each element.  The GBS sample is ordered by $\mathrm{T_{eff}}$.}
    \label{fig: line_by_line_0}
\end{figure*}

\subsubsection{Line-by-Line comparison} \label{line_by_line_results}

To complement the global abundance comparison, we performed a line-by-line analysis of individual spectral lines. In {\tt iSpec}, one can derive the abundances of individual lines separately and calculate a mean of all these measurements at the end. Thus, the global abundances represent the mean abundance derived from all available lines for a given element.
In contrast, the line-by-line analysis involves a detailed comparison of abundances derived from the same individual lines using both methods. This approach enables a direct, one-to-one comparison between EW and spectral synthesis.

Figure \ref{fig: line_by_line_0} shows line-by-line abundance differences between synthesis and EWs ($\Delta A(X) = A(X)_\mathrm{synth} - A(X)_\mathrm{EW}$) for the same elements and in the same color scheme as in Fig.~\ref{GBSV3-R190}. A summary of the mean differences for each element is presented in Table \ref{tab:mean_abs_linebyline_synth_minus_ew_fmt} and the full set of results is provided in online tables. The uncertainties correspond to the combined standard deviation of the mean differences from both methods at each resolution.

The elements Si\,I, Ca\,I, Sc\,II, Ti\,I, Cr\,I, Fe\,II, and Ni\,I show good agreement between the synthesis and EW methods in the line-by-line comparison, with mean differences below 0.08 dex at R190. The trend with effective temperature seen in the global comparison (e.g. Si\,I and Sc\,II, Figure \ref{fig: teff_logg_feh_all}) is significantly reduced  here. This is due to the REW selection criterion, which restricts the analysis to cleaner and more reliable lines. For the remaining elements, the behavior is broadly consistent with the trends identified in the previous section.

Table~\ref{tab:mean_abs_linebyline_synth_minus_ew_fmt} reveals an increase in dispersion toward lower spectral resolution. For example, for Si\,I we find mean differences of 0.032 dex at R190, 0.086 dex at R42, 0.071 dex at R28, and 0.080 dex at R190–R28, with the degraded sample showing a behavior similar to that at R42.

Some elements, such as Ti II and Fe I, still show a noticeable trend in the difference toward the coolest stars. This is consistent with the behavior observed in Figure \ref{fig: abun_1}. In addition, elements affected by HFS, such as V I, Mn I, and Co I, show systematic offsets, although these are generally smaller than in the global comparison of the previous section. For instance, Mn I has a mean difference of 0.7 dex at R190, which is about 0.1 dex lower than the value obtained from the global abundance comparison, although it still represents a large discrepancy. 

The impact of rotational broadening becomes evident for the fastest rotators in our sample. In Fig.~\ref{fig: line_by_line_0}, $\eta$~Boo shows larger abundance differences compared to stars at similar $\mathrm{T_{eff}}$ for all spectral resolutions. Its $v\sin i$ of $ \sim 12.7\,\mathrm{km\,s^{-1}}$ broadens spectral lines and increases blending, thereby affecting the reliability of EW measurements and syntheses. A similar, although less pronounced, effect is observed for HD 49933, which has a lower $v\sin i$ than $\eta$~Boo. The elements Ti\,II and Cr\,I are the most affected.

Overall, the line-by-line approach reduces the discrepancies observed in the global comparison, with mean differences typically decreasing by $\sim$0.2 dex, although residual trends remain for cool stars and HFS-affected elements. Unlike the global comparison, the line-by-line analysis uses the same set of lines for both methods. This removes the bias introduced by the REW selection applied only to the EW measurements, enabling a more direct comparison and isolating differences due to the methods themselves.

\section{Implications for Spectroscopic Surveys} \label{Sec:Discussion}

Our comparison of synthesis and EW-based abundances shows that resolution and method influence reliability in specific cases, such as for elements affected by HFS (Mn I, V I, and Co I). In these cases, EW measurements tend to overestimate abundances, with the effect becoming more pronounced toward the cool regime, where line blending and complex profiles are common. This effect is also obtained in metal-poor stars with intrinsically weak lines. Even at high resolution, HFS-affected lines require synthesis to achieve unbiased results (or proper corrections).

These findings are directly relevant to ongoing and upcoming spectroscopic surveys that aim to map the Milky Way's chemical structure and Galactic archaeology. Inaccurate abundances, especially for $\alpha$-elements, which are tracers of star formation histories, can bias age estimates and population assignment. This  can potentially impact conclusions about the formation and evolution of the Milky Way. For example, there is active research regarding empirical age relations based on chemical abundances, such as the chemical clocks of [Ba/Mg] \citep[see][and references therein]{2024A&A...687A.164V} or more generally chemical ages (\citealp{2023A&A...678A.158A}; \citealp{2024tkas.confE..29B}).
Understanding how methodology and resolution affect abundances is thus essential to maximize the scientific output of on-going spectroscopic surveys.

    {4MOST} (\citealp{2012SPIE.8446E..0TD}) will operate with a resolution mode of $R \sim 20\,000$, slightly below the R28 dataset.
    Our results suggest that robust abundances can be obtained at this resolution, particularly for elements such as Fe I, Ni I, and Si I, where synthesis and EW methods yield consistent values. For cool stars and species affected by HFS or line blending spectral synthesis provides more stable results and should be preferred when possible. We also note that 4MOST presents a similar spectral gap (between 456 and 587\,nm), which may introduce biases in the determination of $\log g$ and certain abundances due to the reduced number of available lines within this wavelength range. 

    {WEAVE} (\citealp{2024MNRAS.530.2688J}) covers a similar resolution range ($R \sim 5000-20\,000$) to 4MOST. The survey's advance data-processing pipeline, combining standardized reduction, calibration, and automated parameter determination, ensures internal homogeneity. Our analysis indicates that at these resolutions, the choice of method will mainly affect elements with complex lines profiles, while others remain robust. For its resolution configurations, EW-based abundances are expected to be less accurate.

    {DESI} (\citealp{2025arXiv250514787K}), operating at $R \sim 2000$-$5000$, lies below the range explored in this work. The dedicated pipelines used to measure APs and radial velocities, based on forward-modeling, ensure homogeneity across the spectra. Nevertheless, our results suggest that for abundances, synthetic-spectrum approaches are the most suitable for chemical abundance analysis at these resolutions.

    {SDSS-V}'s Milky Way Mapper survey combines BOSS low-resolution ($R \sim 2000$) and APOGEE medium-resolution ($R \sim 22\,500$) spectra (\citealp{2025arXiv250707093S}). Our findings reinforce the importance of synthesis-based analyses at these resolutions to ensure consistent abundances for FGK-type stars. The survey’s standardized calibration procedures and analysis frameworks (e.g., \texttt{Astra}) and the abundances presented in \cite{2025AJ....170...96M} are well aligned with this approach. 

    {GALAH DR4} operates at a spectral resolution of $R \sim 28\,000$ (\citealp{2025PASA...42...51B}). The survey adopts a hybrid forward-modeling approach based on synthetic spectra, where APs and chemical abundances are determined simultaneously using neural-network interpolation of synthetic grids computed with Spectroscopy Made Easy (SME; \citealp{1996A&AS..118..595V}; \citealp{2017A&A...597A..16P}), including NLTE effects for several elements.
    
Our study provides an empirical framework to evaluate how spectral resolution and analysis methodology influence abundance precision. These results can aid in defining the optimal analysis strategies for current and future surveys, balancing observing efficiency, spectral coverage, and chemical accuracy, while helping to maintain homogeneity across different datasets and resolutions and ensuring accurate interpretations of Galactic chemical evolution.

\section{Summary and conclusion}\label{Sec:Summary}

Large spectroscopic surveys require robust calibrators to ensure consistent APs and chemical abundances across instruments and resolutions. With their well-determined fundamental parameters, the GBS provide an ideal zero-point reference for this purpose.

In this study, we analyzed 30 southern GBS observed at three different resolutions (R190, R42, R28, and R190-R28) to assess how spectral resolution and analysis method affect the determination of APs and elemental abundances. We constructed a new master linelist based on the GES flags but refined by direct visual inspection at the highest available resolution, retaining only clean, unblended lines for both metal-rich and metal-poor stars.

Using \texttt{iSpec} spectral synthesis, we measured $\mathrm{T_{eff}}$, $\log g$, and $\mathrm{[M/H]}$ across all resolutions (see Figure \ref{fig: teff_logg_feh_all}). We identified a set of free parameters optimized at R190 ($\mathrm{T_{eff}}$, $\log g$, $\mathrm{[M/H]}$, $v_\mathrm{mic}$, $v_\mathrm{mac}$ and $v\mathrm{\sin i}$) and verified that it produces equally consistent results at R42. This demonstrates that resolution around R42 can recover reliable parameters for most FGK stars without the need for ultra-high resolution data. Cool giants and metal-poor stars remain the most challenging cases, showing larger dispersions due to molecular features, weak lines, and NLTE effects, not included here. 

Although the dispersion is higher for R28, the parameters are still consistent with the GBSv3 values. We still identify a systematic offset in $\log g$. To unravel whether this offset arises from spectral resolution or from the limited wavelength coverage, we degraded the R190 spectra to R28 (R190–R28). In this case, the offset disappears, indicating that it is primarily driven by line losses due to spectral gaps rather than by resolution effects alone.

We compared the chemical abundances derived in this work with the MOOG results from Paper VIII, adopting the same atmospheric models and solar abundance scale. We find overall good agreement, particularly for the solar-type stars and for elements such as Mg, Si, Fe, Co, and Ni, with mean differences of 0.01–0.02 dex at R190. Larger discrepancies appear towards cooler temperatures, especially for Ti, V, and Mn. While results at R190 and R42 are consistent, the R28 spectra show larger scatter and more outliers, particularly in metal-poor stars. The comparison with degraded R190-R28 spectra reveals differences of only 0.01 dex, indicating that these differences are mainly driven by line-selection effects rather than spectral resolution itself.

We also derived abundances using both spectral synthesis and EW methodology. At R190, both methods agree well for most elements in solar-type stars; however, significant discrepancies arise towards cool stars. We also found a dependence on line selection, with greater scatter at R28 (e.g., 0.41 dex for Ca I). We concluded that spectral synthesis provides more robust results, particularly for elements affected by HFS (e.g., V I), where EW measurements show systematic offsets (up to 0.62 dex). 

Finally, we performed a line-by-line comparison between spectral synthesis and EW abundances to assess their consistency. Several elements (e.g., Si, Ca, Sc, Ti, Cr, Fe II, and Ni) show improved agreement compared to the previous comparisons, although an increase in dispersion toward lower resolution remains (e.g., Si I from 0.032 dex at R190 to 0.08 dex at lower resolutions).Trends toward cooler stars persisted for some species (e.g., Ti II and Fe I), and elements affected by HFS (e.g., V I, Mn I, Co I) showed systematic offsets, particularly Mn I (0.7 dex),  V I (up to 0.62 dex), while Co I displays smaller deviations. Overall, discrepancies are reduced compared to the global abundances (by 0.1 dex), confirming that line selection mitigates, but does not fully remove, the limitations of the EW method.

Altogether, our results demonstrate that, when using line selection crafted for specific cases, the role of spectral resolution is less critical, although improvements are needed in cooler stars and metal-poor stars, particularly through the inclusion of molecular features and the use of NLTE models. Our findings provide a quantitative framework for survey designers and pipeline developers to anticipate biases and choose the appropriate methodology for their resolution and science goals. By benchmarking different resolutions and methods against the GBS, we demonstrate how to minimize systematic biases in the chemical-abundance catalogs of current and future surveys, thereby improving our understanding of the Milky Way’s formation and evolution.

\section{Data availability}

The data products presented in this paper are available via the CDS. We provide a set of 5 tables summarizing the results of our analysis.  The first table contains the APs ($\mathrm{T_{eff}}$, $\log g$, and [M/H]) derived at different spectral resolutions and for different combinations of free parameters for each star (see Sect.~\ref{stellar_parameters_data}). 
The second, third and fourth tables present the individual comparisons illustrated in Figs.~\ref{GBSV3-R190}, \ref{fig: abun_1}, and \ref{fig: line_by_line_0} whose means are shown in Appendix~\ref{app:means}. The fifth table contains the selected lines for metal-rich and metal-poor regimes.

\phantomsection
\begin{acknowledgements}
I.H.A, C.A.G., and P.J. acknowledge financial support from FONDECYT Regular 1231057. C.A.G. acknowledges support from Agencia Nacional de Investigación y Desarrollo (ANID) through FONDECYT Iniciación 11230741 and FONDECYT Regular 1262342. CS acknowledges support from Observatoire Aquitain des Sciences de l'Univers (OASU). U.H. acknowledges support from the Swedish National Space Agency (SNSA/Rymdstyrelsen). S.B This research has made use of the Astrophysics Data System, funded by NASA under Cooperative Agreement 80NSSC25M7105.
\end{acknowledgements} 

\phantomsection

\bibliographystyle{aa}
\bibliography{tesis_astro}

\clearpage
 
\appendix
\onecolumn

\section{Instruments used for the R42 dataset}
\begin{figure}[h]
\centering
\includegraphics[width=0.5\columnwidth]{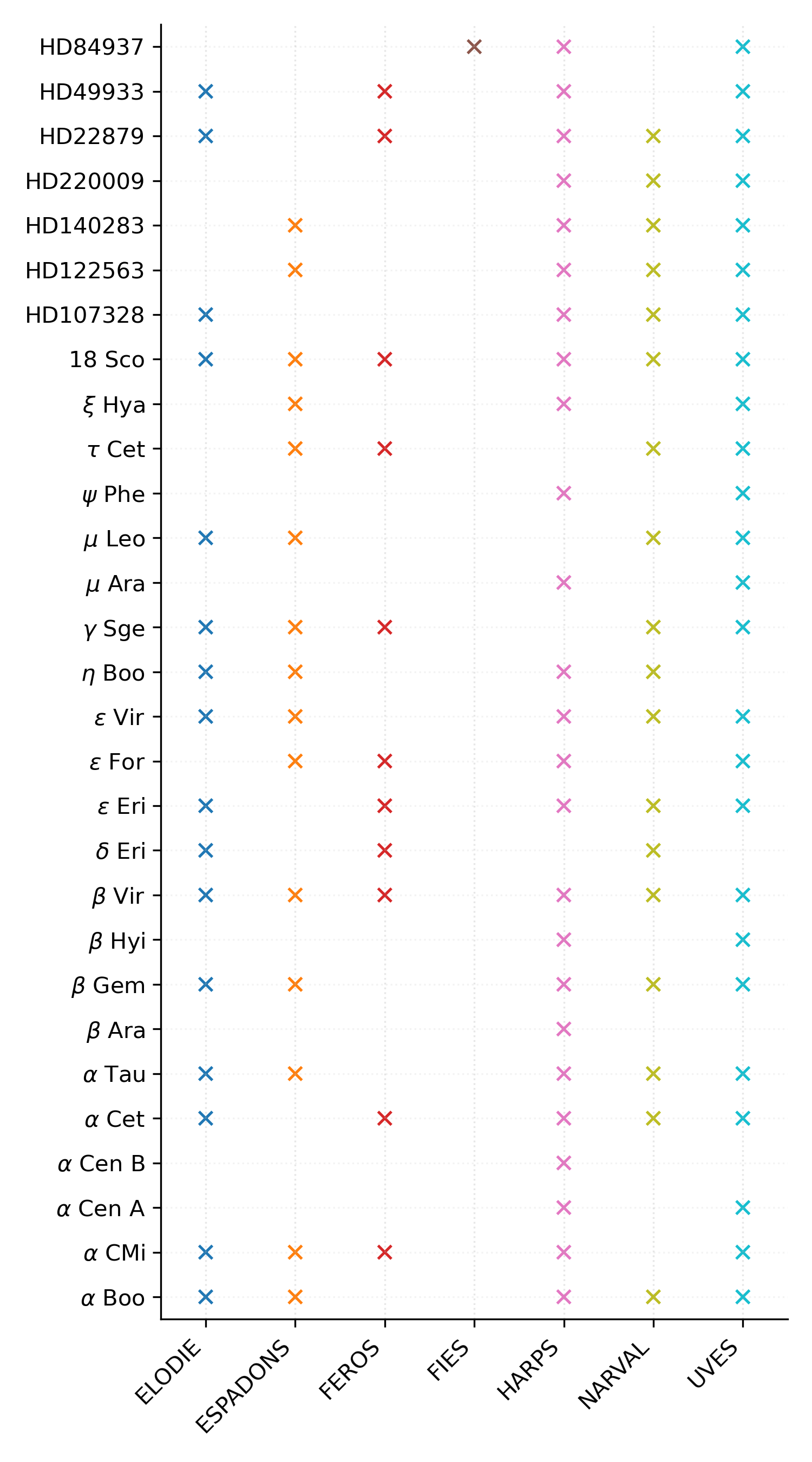}
\caption{Instrumental coverage of the GBS analysed in this work at R42. Crosses indicate the availability of high-resolution spectra for each star–instrument combination.}
\label{R42-STAR-INSTRUMENT}
\end{figure}

In Fig.~\ref{R42-STAR-INSTRUMENT} we show the original instrument for each star used in this study. This is a summary focused on this work but more information about the processing of the data can be found in Paper II and Paper VIII. 

\onecolumn

\section{Results on APs} \label{app: APs}

Table \ref{tabla:std_mean} summarizes the mean differences and standard deviations between the APs derived in this work and the GBSv3 reference values, for $\mathrm{T_{eff}}$, $\log g$, and $\mathrm{[M/H]}$ at resolutions R190, R42, R28, and for the degraded R190-R28 dataset. These results complement Fig. \ref{fig: teff_logg_feh_all} by quantifying the offsets and dispersion introduced by resolution and fitting free parameters.

When using the free parameter set optimized at R190, the mean offsets and dispersion in $\mathrm{T_{eff}}$ and $\log g$ generally increase as the resolution decreases. However, the degraded R190-R28 spectra show smaller offsets and dispersion than the observed R28 dataset, highlighting the additional effect of limited wavelength coverage in the latter.
Allowing the resolution to vary as a free broadening parameter improves the agreement at intermediate resolution, particularly for R42, bringing the results closer to those obtained at R190. For the observed R28 spectra, an offset remains, consistent with the behavior discussed in the main text, but this offset is largely mitigated when using the degraded R190-R28 spectra, confirming that wavelength coverage plays a significant role.
For metallicity, after fixing $\mathrm{T_{eff}}$ and $\log g$ to the GBSv3 values, the dispersion increases toward lower resolution, but the mean differences remain small, indicating that the derived $\mathrm{[M/H]}$ values are overall consistent with the benchmark scale within the uncertainties.

At R190, the derived $v \sin i$ values are in good agreement with the literature (Table \ref{Tabla_GBS}), indicating that the rotational broadening is well resolved at this resolution. However, the situation changes at R28. At this lower resolution, the instrumental broadening becomes comparable to the intrinsic rotational broadening of the fastest rotators in the sample.

In this regime, {\tt iSpec} treats $v \sin i$ simply as an additional broadening parameter in the fit. Therefore, the values derived at R28 should not be interpreted as direct measurements of stellar rotation, but rather as effective broadening parameters that compensate for the combined effects of instrumental resolution and other line-broadening mechanisms. 
This explains why the $v \sin i$ values at R28 differ from the literature, while those at R190 remain consistent.

\begin{table*}[h]
\centering
\setlength{\tabcolsep}{3mm}
\renewcommand{\arraystretch}{1.15}
\footnotesize
\begin{tabular}{l c l r r}
\toprule
Combination & Resolution (R) & Stellar Parameter & Mean & Std. Dev. \\
\midrule

$\mathrm{T_{eff}}$, $\log g$, $\mathrm{[M/H]}$ , vmic, vmac, vsini 
& R190 & Teff  & 74 K   & $\pm$ 77 K \\
&       & $\log g$  & 0.13 dex & $\pm$ 0.10 dex \\
\cmidrule(lr){2-5}
& R42  & Teff  & 118 K & $\pm$ 155 K \\
&       & $\log g$  & 0.20 dex & $\pm$ 0.27 dex \\
\cmidrule(lr){2-5}
& R28  & Teff  & 154 K & $\pm$ 145 K \\
&       & $\log g$  & 0.62 dex & $\pm$ 0.43 dex \\
\cmidrule(lr){2-5}
& R190–R28 & Teff & 85 K & $\pm$ 85 K \\
&           & $\log g$ & 0.19 dex & $\pm$ 0.18 dex \\

\midrule

$\mathrm{T_{eff}}$, $\log g$, $\mathrm{[M/H]}$, vmic, vmac, vsini, R
& R42  & Teff  & 84 K  & $\pm$ 106 K \\
&       & $\log g$  & 0.17 dex & $\pm$ 0.32 dex \\

\midrule

$\mathrm{T_{eff}}$, $\log g$, $\mathrm{[M/H]}$, vmac, R
& R28  & Teff  & 174 K & $\pm$ 151 K \\
&       & $\log g$  & 0.65 dex & $\pm$ 0.51 dex \\
\cmidrule(lr){2-5}
& R190–R28 & Teff & 100 K & $\pm$ 98 K \\
&           & $\log g$ & 0.18 dex & $\pm$ 0.16 dex \\

\midrule

[M/H], vmic, vmac, vsini
& R190 &$\mathrm{[M/H]}$ & 0.03 dex & $\pm$ 0.07 dex \\
\cmidrule(lr){2-5}
& R42  &$\mathrm{[M/H]}$ & 0.10 dex & $\pm$ 0.16 dex \\
\cmidrule(lr){2-5}
& R28  &$\mathrm{[M/H]}$ & 0.16 dex & $\pm$ 0.17 dex \\
\cmidrule(lr){2-5}
& R190–R28 &$\mathrm{[M/H]}$ & 0.05 dex & $\pm$ 0.07 dex \\

\midrule

[M/H], vmic, vmac, vsini, R
& R42 &$\mathrm{[M/H]}$ & 0.10 dex & $\pm$ 0.15 dex \\

\midrule

[M/H], vmac, R
& R28 & $\mathrm{[M/H]}$ & 0.13 dex & $\pm$ 0.10 dex \\
\cmidrule(lr){2-5}
& R190–R28 & $\mathrm{[M/H]}$ & 0.02 dex & $\pm$ 0.07 dex \\

\bottomrule
\end{tabular}

\caption{Difference between our results and GBSv3 for APs determined using different free parameters which are indicated in the column named Combination. }
\label{tabla:std_mean}
\end{table*}

\newpage

\section{Example of spectral lines in cool stars} \label{lines}

\begin{figure*}[h]
    \centering
    \includegraphics[width=1\linewidth]{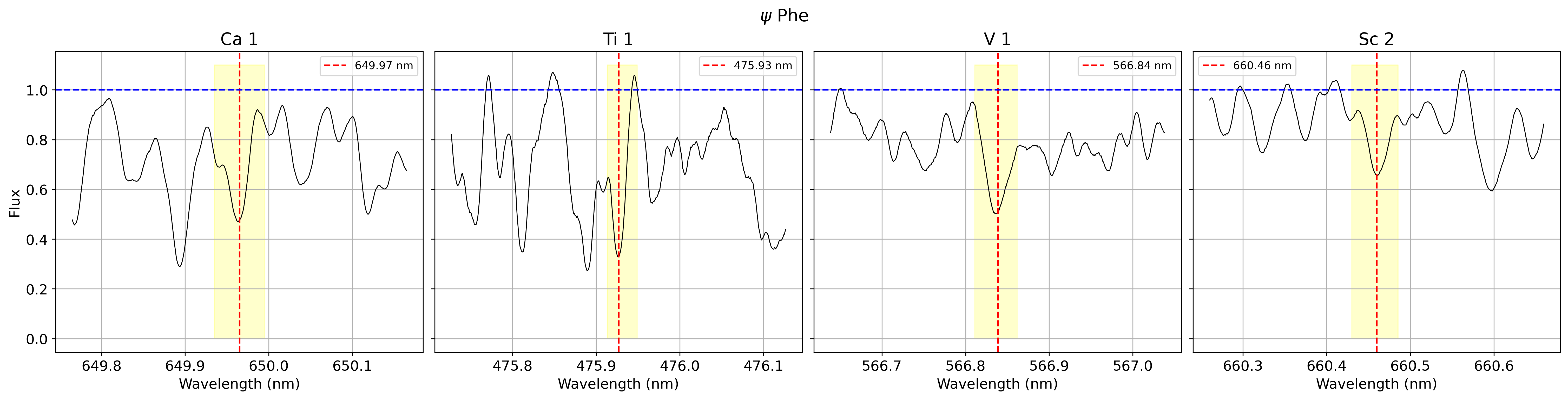}
    \caption{Example lines for $\psi$~Phe at R190. The shaded regions indicate the wavelength intervals used for abundance determination, and the blue dashed line the position of the continuum.}
    \label{fig:psiphe_lines}
\end{figure*}

Figure \ref{fig:psiphe_lines} illustrates representative spectral lines in $\psi$~Phe for selected species (Ca 1, Ti 1, V 1, and Sc 2). These examples highlight cases affected by line blending and, in some regions, by uncertainties in continuum placement due to the crowded spectrum of this cool giant. These effects contribute to the increased scatter and discrepancies discussed in the main text.

\clearpage

\section{Elemental abundance determination and comparison}\label{app:means}
All abundances are determined using the lines listed in Tab.~\ref{tabla:elements}. 
\begin{table}[h]
    \centering
    \setlength{\tabcolsep}{11mm}
    \footnotesize 
    \begin{tabular}{l c c }
        \hline
        Element & $\mathrm{N^\circ}$  Lines &  HFS components \\
        \hline
        Mg\textsuperscript{*} &  3 & N \\
        Si & 4  &  N \\
        Ca & 4  &  N \\
        Sc & 3  &  Y \\
        Ti & 15 & N \\
        V & 6  &  Y \\
        Cr &  4 & N \\
        Mn & 4  &  Y \\
        Fe &  59  & N \\
        Co & 3  &  Y \\        
        Ni & 12  &  N \\
         \hline
        \end{tabular}
        \caption{Number of spectral lines for each chemical element measured and presence of HFS in the lines (Y = yes, N =no). *: elements for which some lines were included in spite of not fulfilling requirements to be considered good lines.}
        \label{tabla:elements}
\end{table}

Table~\ref{tab:mean_abs_delta_gbsv3_fmt} shows the mean difference of the abundance obtained for all stars with our method of global abundances and the MOOG results published as part of Paper VIII. The error represents the standard deviation of the mean. 
\input{tablas_GBSV3/tabla_mediana_GBSV3}

Table~\ref{tab:mean_abs_synth_minus_ew_fmt} shows the mean difference of results obtained with our method using EW and using the global estimation of abundances with synthesis for all stars. We note that for the EW abundances we considered only lines in the $-6.7 < \mathrm{REW} < -4.5$, so the comparison between these two method does not involve the exact same lines. The uncertainties in the values listed correspond to the standard deviation. 
\input{Tablas_ew_synt/tabla_promedio_synth_ew}

\input{tablas_line_by_line/tabla_promedio_linebyline}
Table~\ref{tab:mean_abs_linebyline_synth_minus_ew_fmt} shows the mean differences between the results obtained using EW, as above, and synthesis line-by-line. In this case,  the comparison entails the exact same lines and only shows the difference in methodology.

\twocolumn
%
%
\end{document}

%% file: tablas_GBSV3/tabla_mediana_GBSV3.tex
\begin{table*}[h]
\centering
\caption{Mean absolute abundance differences relative to GBSV3 for each element and resolution, shown as mean $\pm$ standard deviation.}
\label{tab:mean_abs_delta_gbsv3_fmt}
\begin{tabular}{lcccc}
\toprule
Element & R190 & R45 & R28 & R190-R28 \\
\midrule
Ca 1 & 0.016 $\pm$ 0.028 & 0.012 $\pm$ 0.020 & 0.049 $\pm$ 0.052 & 0.018 $\pm$ 0.033 \\
Co 1 & 0.012 $\pm$ 0.023 & 0.015 $\pm$ 0.025 &  & 0.011 $\pm$ 0.022 \\
Cr 1 & 0.024 $\pm$ 0.045 & 0.026 $\pm$ 0.049 &  & 0.032 $\pm$ 0.063 \\
Fe 1 & 0.009 $\pm$ 0.013 & 0.011 $\pm$ 0.014 & 0.024 $\pm$ 0.027 & 0.013 $\pm$ 0.017 \\
Fe 2 & 0.011 $\pm$ 0.013 & 0.015 $\pm$ 0.019 &  & 0.018 $\pm$ 0.021 \\
Mg 1 & 0.017 $\pm$ 0.013 & 0.021 $\pm$ 0.018 &  & 0.020 $\pm$ 0.019 \\
Mn 1 & 0.022 $\pm$ 0.038 & 0.025 $\pm$ 0.038 &  & 0.024 $\pm$ 0.040 \\
Ni 1 & 0.013 $\pm$ 0.016 & 0.014 $\pm$ 0.018 & 0.030 $\pm$ 0.026 & 0.014 $\pm$ 0.017 \\
Sc 2 & 0.017 $\pm$ 0.016 & 0.018 $\pm$ 0.019 & 0.068 $\pm$ 0.081 & 0.035 $\pm$ 0.026 \\
Si 1 & 0.010 $\pm$ 0.016 & 0.012 $\pm$ 0.021 & 0.015 $\pm$ 0.017 & 0.007 $\pm$ 0.008 \\
Ti 1 & 0.024 $\pm$ 0.035 & 0.020 $\pm$ 0.026 & 0.042 $\pm$ 0.053 & 0.027 $\pm$ 0.041 \\
Ti 2 & 0.030 $\pm$ 0.055 & 0.032 $\pm$ 0.058 & 0.077 $\pm$ 0.087 & 0.043 $\pm$ 0.078 \\
V 1 & 0.039 $\pm$ 0.062 & 0.071 $\pm$ 0.205 & 0.053 $\pm$ 0.061 & 0.047 $\pm$ 0.074 \\
\bottomrule
\end{tabular}
\end{table*}

%% file: Tablas_ew_synt/tabla_promedio_synth_ew.tex
\begin{table*}[ht]
\centering

\caption{Mean absolute abundance differences between synthetic-spectrum and EWs methods for each element and resolution, shown as mean $\pm$ standard deviation.}
\label{tab:mean_abs_synth_minus_ew_fmt}
\begin{tabular}{lcccc}
\toprule
Element &  R190 & R42 & R28 & R190→R28 \\
\midrule
\midrule
Ca 1 & 0.075 $\pm$ 0.045 & 0.097 $\pm$ 0.241 & 0.414 $\pm$ 0.368 & 0.039 $\pm$ 0.046 \\
Co 1 & 0.272 $\pm$ 0.206 & 0.291 $\pm$ 0.295 &  & 0.336 $\pm$ 0.255 \\
Cr 1 & 0.073 $\pm$ 0.116 & 0.092 $\pm$ 0.162 &  & 0.144 $\pm$ 0.173 \\
Fe 1 & 0.076 $\pm$ 0.151 & 0.188 $\pm$ 0.303 & 0.277 $\pm$ 0.208 & 0.129 $\pm$ 0.169 \\
Fe 2 & 0.146 $\pm$ 0.388 & 0.245 $\pm$ 0.448 & 0.573 $\pm$ 0.632 & 0.245 $\pm$ 0.337 \\
Mn 1 & 0.800 $\pm$ 0.548 & 0.739 $\pm$ 0.458 & 1.142 $\pm$ 0.473 & 0.671 $\pm$ 0.293 \\
Ni 1 & 0.075 $\pm$ 0.081 & 0.206 $\pm$ 0.330 & 0.210 $\pm$ 0.178 & 0.156 $\pm$ 0.132 \\
Sc 2 & 0.132 $\pm$ 0.227 & 0.173 $\pm$ 0.211 & 0.395 $\pm$ 0.290 & 0.250 $\pm$ 0.301 \\
Si 1 & 0.114 $\pm$ 0.207 & 0.255 $\pm$ 0.275 & 0.226 $\pm$ 0.190 & 0.162 $\pm$ 0.188 \\
Ti 1 & 0.150 $\pm$ 0.255 & 0.191 $\pm$ 0.461 & 0.269 $\pm$ 0.254 & 0.159 $\pm$ 0.112 \\
Ti 2 & 0.140 $\pm$ 0.306 & 0.345 $\pm$ 0.775 & 0.646 $\pm$ 0.636 & 0.237 $\pm$ 0.389 \\
V 1 & 0.620 $\pm$ 0.923 & 0.389 $\pm$ 0.384 & 0.321 $\pm$ 0.265 & 0.459 $\pm$ 0.647 \\
\bottomrule
\end{tabular}
\end{table*}

%% file: tablas_line_by_line/tabla_promedio_linebyline.tex
\begin{table*}[ht]
\centering
\caption{Mean absolute abundance differences between synthetic-spectrum and EWs methods for each element and resolution, shown as mean $\pm$ standard deviation.}
\label{tab:mean_abs_linebyline_synth_minus_ew_fmt}
\begin{tabular}{lcccc}
\toprule
Element & R190 & R42 & R28 & R190-R28 \\
\midrule
Ca 1 & 0.040 $\pm$ 0.041 & 0.058 $\pm$ 0.081 & 0.160 $\pm$ 0.183 & 0.035 $\pm$ 0.060 \\
Co 1 & 0.164 $\pm$ 0.103 & 0.188 $\pm$ 0.140 &  & 0.242 $\pm$ 0.215 \\
Cr 1 & 0.048 $\pm$ 0.054 & 0.085 $\pm$ 0.117 &  & 0.059 $\pm$ 0.041 \\
Fe 1 & 0.084 $\pm$ 0.150 & 0.085 $\pm$ 0.126 & 0.236 $\pm$ 0.253 & 0.125 $\pm$ 0.176 \\
Fe 2 & 0.082 $\pm$ 0.175 & 0.124 $\pm$ 0.162 & 0.225 $\pm$ 0.020 & 0.142 $\pm$ 0.195 \\
Mn 1 & 0.700 $\pm$ 0.523 & 0.508 $\pm$ 0.266 & 0.322 $\pm$ 0.209 & 0.660 $\pm$ 0.410 \\
Ni 1 & 0.081 $\pm$ 0.122 & 0.115 $\pm$ 0.159 & 0.084 $\pm$ 0.085 & 0.117 $\pm$ 0.099 \\
Sc 2 & 0.072 $\pm$ 0.054 & 0.114 $\pm$ 0.070 & 0.112 $\pm$ 0.123 & 0.130 $\pm$ 0.132 \\
Si 1 & 0.032 $\pm$ 0.033 & 0.086 $\pm$ 0.074 & 0.071 $\pm$ 0.076 & 0.080 $\pm$ 0.112 \\
Ti 1 & 0.020 $\pm$ 0.021 & 0.084 $\pm$ 0.113 & 0.089 $\pm$ 0.113 & 0.049 $\pm$ 0.045 \\
Ti 2 & 0.058 $\pm$ 0.158 & 0.135 $\pm$ 0.196 & 0.524 $\pm$ 0.406 & 0.129 $\pm$ 0.148 \\
V 1 & 0.258 $\pm$ 0.265 & 0.288 $\pm$ 0.240 & 0.295 $\pm$ 0.249 & 0.257 $\pm$ 0.212 \\
\bottomrule
\end{tabular}
\end{table*}